\documentclass[aps,prb,twocolumn,showpacs,superscriptaddress,groupedaddress]{revtex4}
\usepackage{graphicx}  
\usepackage{dcolumn}   
\usepackage{bm}        
\usepackage{amssymb}   
\usepackage{physics}
\usepackage{youngtab}
\usepackage{array}
\usepackage{verbatim}
\usepackage{float}

\usepackage[utf8]{inputenc}

\begin{document}
\title{PHASE TRANSITION IN $SU(N)\times U(1)$ GAUGE THEORY WITH MANY FUNDAMENTAL BOSONS}
\author{Ankur Das\\
University of Kentucky \\ email \href{mailto:ada258@g.uky.edu}{ada258@g.uky.edu} }
\date{\today}

\begin{abstract}
Here we study the Renormalization group flow of $SU(N)\times U(1)$ gauge theory with $M$-fundamental bosons in $4-\epsilon$ dimension by calculating the beta functions. We found a new stable fixed point in the zero mass plane for $M>M_\text{crit}$ by expanding upto $O(\epsilon)$. This indicates a second order phase transition. We also calculated the critical exponents in both $\epsilon$ expansion and also in the large-$M$ expansion.

\end{abstract}

\maketitle
\section{Introduction}
Phase transitions in gauge theories are very interesting because gauge theories appear as effective theories in many physical problems. Historically, in particle physics gauge theories have been studied in detail because of their potential application to phenomenology. More recently, there are several examples of emergent gauge degrees of freedom in condensed matter physics\cite{sachdev,senthil,senthil2,lammert,lee,huo,blasone}. Phase transitions in those theories hold very rich physics. We will be concerned solely with continuous gauge symmetries.

\begin{figure}
    \centering
    \includegraphics[scale=0.2]{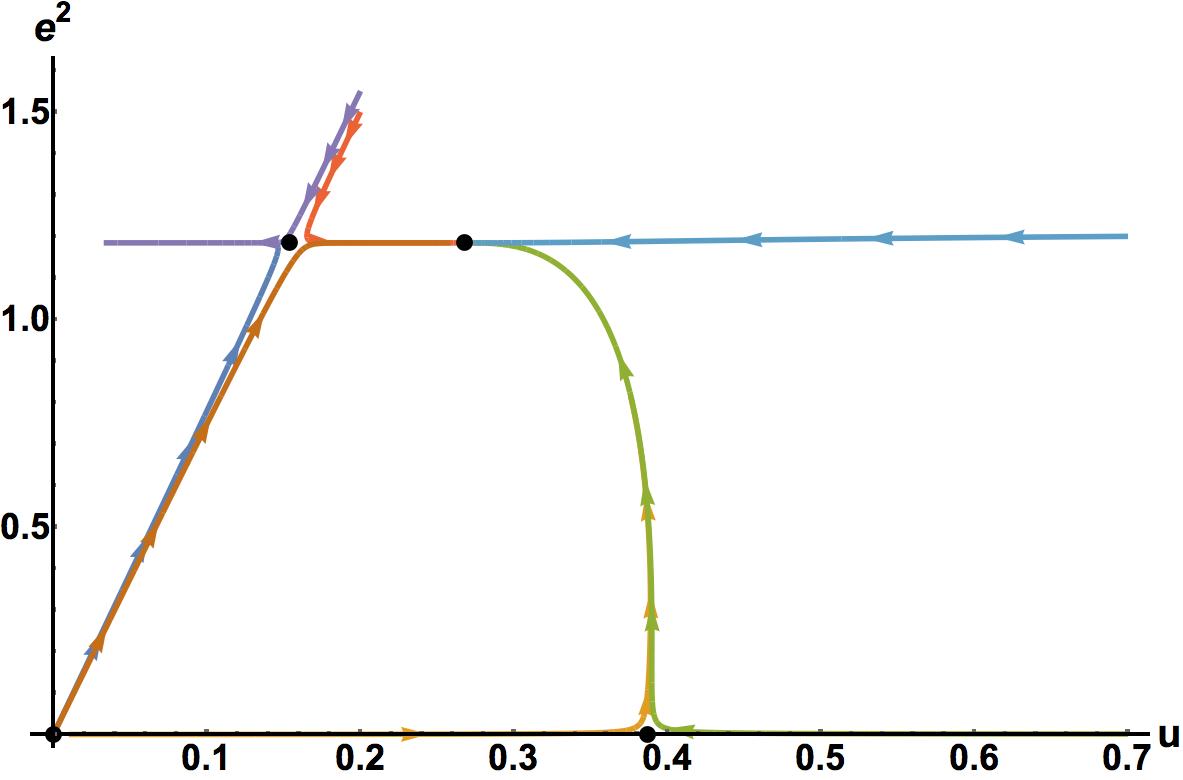}
    \caption{Flow diagram in the $u-e^2(=\alpha)$ plane for $M>M_\text{cric}$. As one can see there are $4$ fixed points(The fixed points are also plotted in black dots). One can see the Gaussian fixed point and the well known and famous WF fixed point. But there are two new charged fixed point there which are present only for $M>M_\text{cric}$. One of them is a stable fixed point. There exist also a charged fixed point which is not stable in this plane. This is what was found by Halperin-Lubansky-Ma\cite{HaLuMa}.}
    \label{flow2d}
\end{figure}

The simplest example of a phase transition in a continuous gauge theory is in $U(1)$ gauge theory with a single boson. This is the Ginzburg-Landau theory of superconductor-insulator transition\cite{gor}. Fluctuations around mean field were first studied by Coleman and Weinberg\cite{weinberg}, who found that in $d=4$ the theory undergoes a first-order phase transition. This conclusion was verified independently by Halperin, Lubensky and Ma(HLM)\cite{HaLuMa}, who also carried out an $\epsilon$ expansion in $d=4-\epsilon$ dimensions to first order in $\epsilon$. They also showed $d=3$ by integrating out the gauge degrees of freedom that the transition becomes weakly first order. Generalizing to $M$ complex boson fields they found for $M>M_\text{crit}=182.95$ two more fixed points appear, as shown in Fig. \ref{flow2d}. It is seen that for $M>M_\text{crit}$ there is a stable fixed point in the zero mass plane indicating a second order phase transition. Halperin, Lubensky, and Ma also calculated the critical exponents for the transition in the $\epsilon$ expansion and in fixed dimension $d=3$ in the large-$M$ approximation.

The case of an $SU(2)$ gauge field coupled to $M$ fundamental bosons has been studied more recently by Arnold and Yaffe.\cite{arnold}. They found a picture very similar Fig. \ref{flow2d} in the $\epsilon$ expansion. To $O(\epsilon)$ they found that  for $M>M_\text{crit}=359$ there are two charged fixed points. One of them is attractive in the $b-g^2$ plane, again indicating a second-order phase transition. The $SU(2)\times U(1)$ case is known as the electroweak phase transition.

It is known from several numerical studies\cite{laine,laine2,laine3} in lattice gauge theory that in the case of $M=1$ there exists a critical ratio of the couplings such that for $b/g^2>C$ there is no phase transition at all and for $b/g^2<C$ the transition is first order. The second order phase transition exists only if $b/g^2=C$. The reason is that for $b/g^2>C$ no symmetry is broken in the $SU(2)$ transition.

But this picture changes in a very significant way when more than one species/flavours of boson are introduced (these transform as higher representations under the gauge group). In that case as Fradkin and Shenker\cite{fradkin} show in lattice gauge theory, a phase transition does occur for all the values of ratio of couplings. In a gauge theory with a non-trivial center, the center survives for higher representations in unitary gauge if the boson is in the adjoint representation. Introducing $M$ species of bosons leads to  a global $U(M)$ symmetry\cite{arnold2}. In the unitary gauge the $SU(N)$ gauge symmetry breaks down but this $U(M)$ symmetry survives. The phase transition corresponds to spontaneous breaking of this $U(M)$ symmetry.

In this paper we study $SU(N)\times U(1)$ theory with $M$ flavours of bosons. Such a theory arises in a completely different context, the study of $SU(M)$ antiferromagnets on a square lattice\cite{sachdev}.

The Hamiltonian of this model is,
\begin{equation}
    \mathcal{H}=\frac{J}{M}\sum_{\langle i,j\rangle} {\hat{S}^\beta}_\alpha(i){\hat{S}^\alpha}_\beta(j)
\end{equation}

\begin{figure}[h]
    \centering
    \includegraphics[scale=0.3]{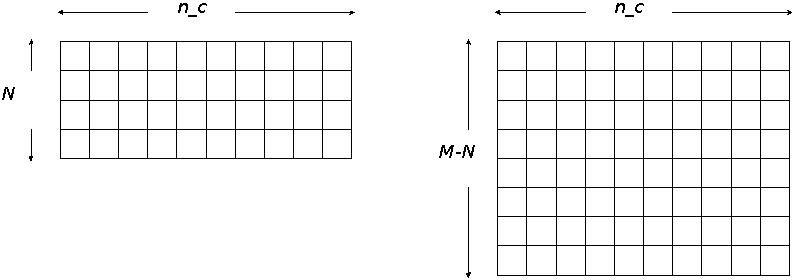}
    \caption{The representation in terms of Young Tableau of $SU(M)$ Lie group of the spins on sub-lattice $A$ and $B$. The number of boxes in every row is $n_c$ where the number of boxes in the column for the $A$ sub-lattice is $N$ and $M-N$ for $B$ sub-lattice}
    \label{tabl}
\end{figure}

Where ${\hat{S}^\beta}_\alpha(i)$ are the generators of $SU(M)$ and $\langle i,j\rangle$ represents nearest neighbour sum on this bipartite square lattice. The representation of the spins sitting in two sublattices ($A$ and $B$) can be described using the two integers describing the Young tableau, $n_c$ and $M$. The representation of the spins are described in Fig. \ref{tabl}. For the $A$ sublattice the number of boxes in the column of the young tableau is $N$ where for the $B$ sub-lattice the boxes in the column is $M-N$. The number of boxes in every row is fixed to be $n_c$.

Now, we introduce boson (Schwinger boson\cite{auerbach,auerbach2}) operators $b^{\alpha a}(i)$ for sublattice $A$ and $\bar{b}_{\alpha a}(j)$ on each sublattice $B$ with the constraint,
\begin{subequations}
    \begin{equation}
        b^\dagger_{\alpha a}(i) b^{\alpha b}(i)=\delta^b_a n_c,\text{   no sum on }i
    \end{equation}
    \begin{equation}
        {\bar{b}^{\alpha a \dagger}}(j)\bar{b}_{\alpha b}(j)=\delta^b_a n_c,\text{   no sum on }j
    \end{equation}
\end{subequations}

And the spin operators will be,
\begin{subequations}
    \begin{equation}
        \hat{S}_\beta^\alpha(i)=\sum_{a=1}^N b^\dagger_{\beta a}(i)b^{\alpha a}(i),~~~ i\in A \text{~sublattice}
    \end{equation}
    \begin{equation}
        \hat{S}^\beta_\alpha(j)=-\sum_{a=1}^N {\bar{b}^{\beta a\dagger}}(j)\bar{b}_{\alpha a}(j),~~~ j\in B \text{~sublattice}
    \end{equation}
\end{subequations}
In the functional integral representation of the partition function of the $\mathcal{H}$ can be written as,

\begin{equation}
    \mathcal{Z}=\int \mathcal{D}Q\mathcal{D}b\mathcal{D}\bar{b}\mathcal{D}\lambda\exp\left[ -\int_0^\beta \mathcal{L}d \tau \right]
\end{equation}

Where,
\begin{multline}
    \mathcal{L}=\sum_{i\in A}\left[ b^\dagger_{\alpha a}(i)\left(\delta^a_b\frac{d}{d\tau}+i\lambda^a_b(i)\right)b^{\alpha b}(i)-i\lambda^a_a(i)n_c \right]\\
    +\sum_{j\in B}\left[ \bar{b}^{\alpha a\dagger(j)}\left(\delta^b_a\frac{d}{d\tau}+i\lambda^b_a(j)\right)\bar{b}_{\beta b}(j)-i\lambda^a_a(j)n_c \right]\\
    +\sum_{i,\hat{n}}\left[ \frac{M}{J}\abs{{Q^a_b}_{i,i+\hat{n}}}^2-\left( {{{Q^{*b}}_a}_{i,i+\hat{n}}b^{\alpha a}(i)\bar{b}_{\alpha b}}(i+\hat{n})+H.c. \right) \right]
\end{multline}
Where $\lambda(i)$ is the lagrange multiplier which fixes the number of bosons per site to be $n_c$. This Lagrangian is local $U(N)$ invariant. The field ${Q^a_b}_{i,i+\hat{n}}$ is the Hubbard-Stratonovich field. Now we can do a mean field approximation of this theory and in that mean field approximation. In which $\lambda$ and $Q$ becomes constant. The fluctuation of this these fields around that mean field value will be,
\begin{subequations}
    \begin{equation}
        Q^a_{b i,i+\hat{n}}=\left[ \bar{Q}\delta^a_c+q^a_c(i) \right]\exp\left[i\hat{n}\cdot B(i)\right]^c_b
    \end{equation}
    \begin{equation}
        i\lambda^a_b(i)=\bar{\lambda}\delta^a_b+i{B^a_b}_\tau(i), \text{for } i\in A
    \end{equation}
    \begin{equation}
        i\lambda^a_b(j)=\bar{\lambda}\delta^a_b-i{B^a_b}_\tau(j), \text{for } j\in B
    \end{equation}
\end{subequations}

Where $B_\mu$ is the $U(N)$ gauge field and $q$ is amplitude fluctuation. One can do a long wavelength approximation to this to get action,
\begin{equation}
    S_\text{eff}=\int d^dr\int_0^{c\beta}d\tilde{\tau}\frac{a^{1-d}}{2\sqrt{d}}\left[ \abs{(\partial_\mu \delta^a_b-i B^a_{\mu b}z^{\alpha b})} +\frac{\Delta^2}{c^2}\abs{z^{\alpha a}}^2\right]
\end{equation}
The gauge field can be broken into a $U(1)$ and a $SU(N)$ part(trace and trace less part),
\begin{equation}
    B^a_{b\mu}=\delta^a_b A_\mu+W^a_{b\mu}
\end{equation}

A gradient expansion of this will give us the $SU(N)\times U(1)$ theory with $M$ flavour of bosons.
\par
The phases $SU(M)$ antiferromagnet are known for $N=1$. We want to check how the order of the phase transition depends on the number of flavor for the $SU(N) \times U(1)$. We want to check this in two ways. First we can try to integrate out the gauge field which we will do for $M=1$ and $N=2$ to show that for a single flavor in fundamental representation there is no second order transition at least for $N=2$. Next, as we want to study the the theory that arises from the $SU(M)$ anti-ferromagnets. We will study the RG flow of this theory for arbitrary $M$ and $N$ and the fixed point structure of the theory.

The phase transition in $SU(M)$ magnets(Heisenberg model) has been studied numerically before. Kawashima and Tanabe\cite{naoki} found evidence of emergent $U(1)$ symmetry of the ground state space of the $SU(M)$ Heisenberg model with the fundamental representation. Beach et al.\cite{beach} developed a quantum Monte Carlo algorithm to simulate this model for continuous $M$ in total singlet basis and found a phase transition between Ne\'el and VBC columnar phase occurring at $M_c=4.57(5)$. They also identified the phase transition to be second order with critical exponents, $z=1$ and $\beta/\nu=0.81(3)$.

\section{Effect of gauge fluctuations}
First we will try to integrate out the gauge field to see what happens for $M=1$(in the Unitary gauge) to the action defined as,
\begin{multline}
    S[\psi,\vec{B},\vec{W}^a]=\int d^3x \bigg[ \abs{(\partial_\mu -i y A_\mu - i g T^a W_\mu^a)\psi}^2 \\
    +\frac{1}{4}F^{\mu\nu}F_{\mu\nu}+\frac{1}{4}G^{a\mu\nu}G^a_{\mu\nu}+a\abs{\psi}^2+\frac{b}{2}\abs{\psi}^4 \bigg]
\end{multline}
Where,
\begin{align}
    F_{\mu\nu}&=\partial_\mu A_\nu-\partial_\nu A_\mu\\
    G^a_{\mu\nu}&=\partial_\mu W^a_\nu-\partial_\nu W^a_\mu+g f_{abc}W^b_\mu W^c_\nu
\end{align}
and as usual,
\begin{equation}
    a=\frac{a'(T-T_c)}{T_c}
\end{equation}

This in pure $U(1)$ case leads to a weak-first order phase transition as the gauge field around mean field approximation of the order parameter picks up a mass(in other words this will give us Meissner effect with a penetration depth defined by the mass).\cite{HaLuMa}

The $\psi$ field has $N$ components. Now, the minimum of this action is when all the fluctuations of fields are zero and $\abs{\psi}=const$. This value of the constant is well known, i.e.
\begin{equation}
    \abs{\psi}=\pm\sqrt{-\frac{a}{b}}
\end{equation}
Now for $N=2$ one can choose a gauge to make, $\psi_1=0$ and $\psi_2=\sqrt{-a/b}$.\par
Now the question comes of the Ginzburg-Criteria which we can find after a few calculations is,
\begin{equation}
    \frac{T-T_c}{T_c}<\frac{1}{8\pi^2}\frac{b^2 T_c^2}{a'}.
\end{equation}
In the case of superconductor theory we actually know the microscopic theory(BCS theory) and from there one can exactly find these coefficients $a,b$ in terms of microscopic parameters.\cite{gor} This ensures that the Ginzburg Criteria is met and we can actually use constant Mean Field solution. In our case we don't know the microscopic theory but for now we will assume that the $\psi$(order parameter) fluctuation is very small and we can use the mean field value of the field.\par
Next we need to consider the case where we choose a specific gauge and want to calculate the effect of the gauge field fluctuations. Again we will do it for $N=2$. We choose the gauge such that $\psi_1=0$ and $\psi_2=v$. For $N=2$ the generators are,
\begin{equation}
    T^a=\frac{1}{2}\sigma^a.
\end{equation}
We find that the mass matrix of the fields is not diagonalized. After the mass matrix diagonalization we find that there will be $3$ gauge fields with mass and one massless gauge field(all U(1)) but interacting with each other. The massive fields are $W^1_\mu,W^2_\mu$ with mass square, $m_1^2=m_2^2=(1/2)g^2v^2$ and $Z_\mu$ with mass square, $m_Z^2=v^2(g^2+4y^2)/2$. We will also have a massless field $A_\mu$. The definition of $A_\mu$ and $z_\mu$ is,
\begin{align}
    A_\mu&=\sin \theta_W b_\mu+\cos \theta_W W^3_\mu\\
    Z_\mu&=\cos \theta_W B_\mu-\sin \theta_W W^3_\mu
\end{align}
where, $\sin \theta_W=g/\sqrt{g^2+4y^2}$ and $\cos \theta_W=2y/\sqrt{g^2+4y^2}$.\par
As mentioned before we want to calculate,
\begin{equation}
    \exp(-S(\psi)/T)=\int\mathcal{D}A\mathcal{D}Z\mathcal{D}W^i\exp[-S[\psi,\vec{B},\vec{W}^a]/T].
\end{equation}
For our gauge we find that,
\begin{equation}\label{action}
    \frac{d S}{d v}=2(vol)a v+2(vol)bv^3+(g^2/4)\langle {W^i_\mu}^2\rangle v+(g^2/4+y^2)\langle Z_\mu^2\rangle v.
\end{equation}
These averages can be calculated to be, for small $v$
\begin{align}
    \langle {W^i_\mu}^2\rangle=\frac{2(vol)}{\pi^2}\left[ \Lambda-\frac{m_i \pi}{2} \right]\\
    \langle {Z_\mu}^2\rangle=\frac{2(vol)}{\pi^2}\left[ \Lambda-\frac{m_Z \pi}{2} \right]
\end{align}

Putting all this to \ref{action} and integrating over $v$ we get,
\begin{equation}
    \frac{S}{vol}=\left[ \left( a+\frac{3\Lambda}{2\pi^2} \right)v^2+\frac{b}{2}v^4-3\left(2 g+ \sqrt{g^2+4y^2}\frac{v^3}{4\sqrt{2}\pi}  \right) \right]
\end{equation}

This introduces a first order phase transition exactly like in $U(1)$-case.\cite{HaLuMa}. From this one can calculate the size of the phase transition etc.

\section{Beta functions and fixed points}
The more general way to find $\beta$-function is to carry out RG calculations in $d=4-\epsilon$ and for general $N$ using dimensional regularization. we define here for simplicity of the calculations $\alpha_1=y^2$ and $\alpha_2=g^2$.\cite{pes,sred,igor}

Thus the beta functions are,

\begin{align}
    \beta_{\alpha_1}&=\epsilon \alpha_1-\frac{\alpha_1^2 N M}{24 \pi^2}\\
    \beta_{\alpha_2}&=\epsilon\alpha_2-\frac{\alpha_2^2}{48N\pi^2}(M-22N)\\
    \beta_a&=a\left[2-\frac{b(N M + 1)}{8\pi^2}+\frac{3\alpha_2}{8\pi^2}\left(\frac{N^2-1}{2N}\right)+\frac{3\alpha_1}{8\pi^2}\right]
\end{align}
\begin{multline}
    \beta_b=\epsilon b-\frac{b^2(N M +4)}{8\pi^2}-\frac{3\alpha_1^2}{4\pi^2}-\frac{3\alpha_2^2(N^3+N^2-4N+2)}{32\pi^2N^2}\\
    -\frac{3\alpha_1\alpha_2}{\pi^2 N}(N-1)+\frac{3b\alpha_1}{4\pi^2}+\frac{3b\alpha_2}{4\pi^2}\left( \frac{N^2-1}{2N} \right).
\end{multline}

\par
One can easily see from the structure of the $\beta$-function that for $N=1$, $\beta_{\alpha_1}, \beta_{a},\beta_b$ completely decouples from the $\alpha_2$ and as one can check that it has the correct structure for $U(1)$ gauge theory with multiple scalar.\cite{HaLuMa,igor} Next one can look into the fixed point structure of this theory. There are $8$ possible fixed points of these $\beta$-functions. Two of them are the old Gaussian and the Wilson-Fisher fixed point and fixed points where there is no $SU(N)$ or $U(1)$ charge.\cite{HaLuMa} As before the $U(1)$-charged fixed points do not exist for $NM<182.952$. There are four more fixed points that arise in the theory and one of them is critical as that one is completely stable in all direction except for the temperature(mass) direction. This point is doubly charged. But this fixed point does not exist for $M<M_\text{crit}$. This $M_\text{crit}$ is different for different values of $N$. For example for $N=2$, $M_\text{crit}=1277.47$. There are two singly charged ($SU(N)$ charge) fixed points also. This $SU(N)$ charged fixed points also have some critical value of $M$ as a function of $N$. As previously calculated for $N=2$ this critical value is $359$.\cite{arnold}
\begin{figure}[!h]
    \centering
    \includegraphics[scale=0.27]{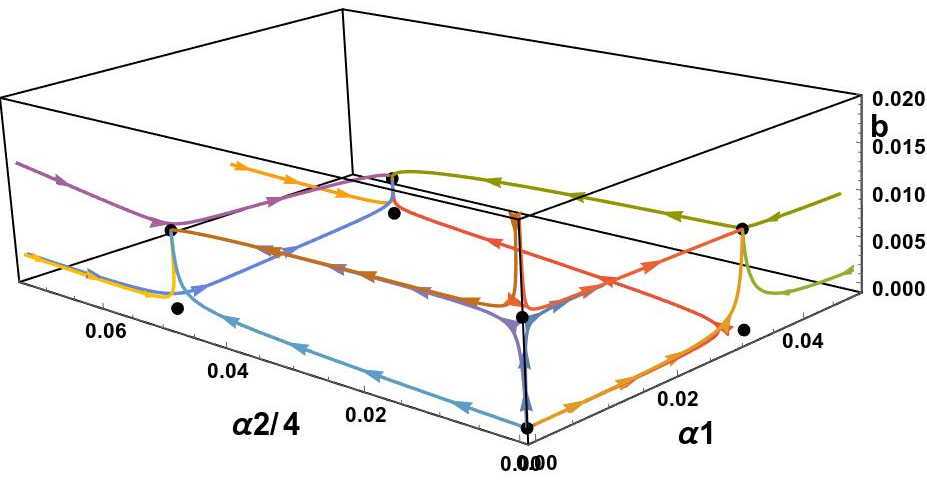}
    \caption{RG flow diagram for the $N=2$ and $M=1500$ where the all attractive point exists. As we can see here there are 8 fixed point and one attractive in all direction(other than mass). That fixed point denotes the second order phase transition of the system}
\end{figure}
\begin{figure}[!h]
    \centering
    \includegraphics[scale=0.25]{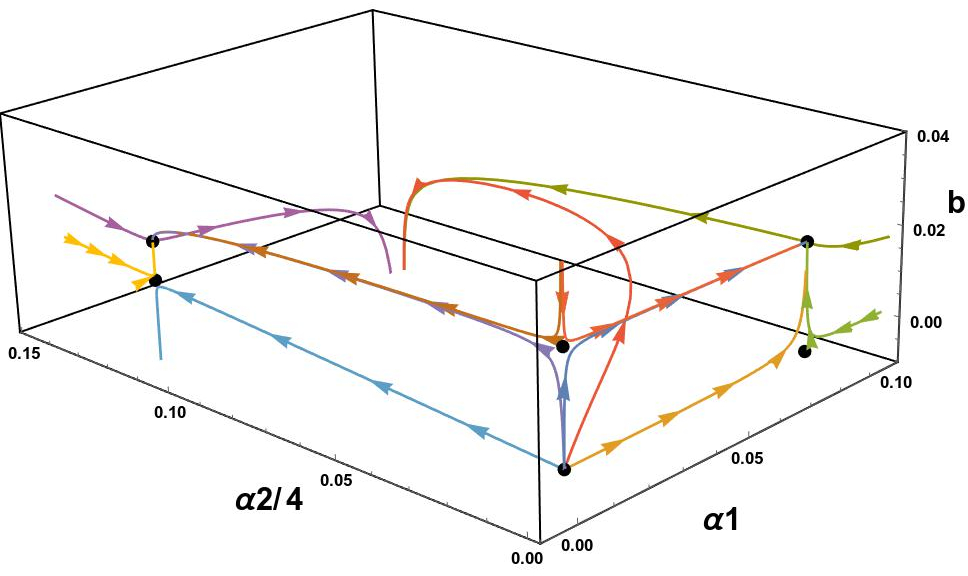}
    \caption{RG flow diagram for the $N=2$ and $M=1100$ where the all attractive point does not exists and as we can see that the flow does not have any more $8$ fixed points. The attractive doubly charged fixed point is now gone and all flow with any non-zero initial charge flows to negative mass denoting a first order phase transition}
\end{figure}
\section{Critical exponents}
The critical exponents of this phase transition can be easily calculated in the regular way and we can see that $\nu\rightarrow 1$ and $\eta\rightarrow 0$ as $M\rightarrow\infty$ for $\epsilon=1$. In terms of fixed point value of the parameters($a^*=0,b^*,\alpha_1^*,\alpha_2^*$)\cite{kiom}
\begin{align}
    \frac{1}{\nu}&=2-\frac{b^*(N+1)}{8\pi^2}+\frac{3\alpha_1^*}{8\pi^2}+\frac{3\alpha_2^*}{8\pi^2}\left( \frac{N^2-1}{2N} \right)\\
    \eta&=-\left[ \frac{3\alpha_1^*}{4\pi^2}+\frac{3\alpha_2^*}{8\pi^2}\left( \frac{N^2-1}{2N} \right) \right]
\end{align}
As we have seen these beta functions has a very interesting structure of fixed points (we have $M>N$). There are $8$ fixed points but not all of them exists at every value of $M$ and $N$. The $M$ and $N$ comes from the microscopic theory.\cite{sachdev}. For $N=1$ the theory contains only the abelian gauge field. The question one needs to ask is for what values of $N$ and $M$ there exist a doubly charged critical point. We can easily find out the relation between $N$ and $M_\text{cric}$. That relation is quadratic,
\begin{equation}
    M_\text{crit}=607.765 + 174.594 N + 106.058 N^2
\end{equation}

\begin{figure}[h]
    \centering
    \includegraphics[scale=0.3]{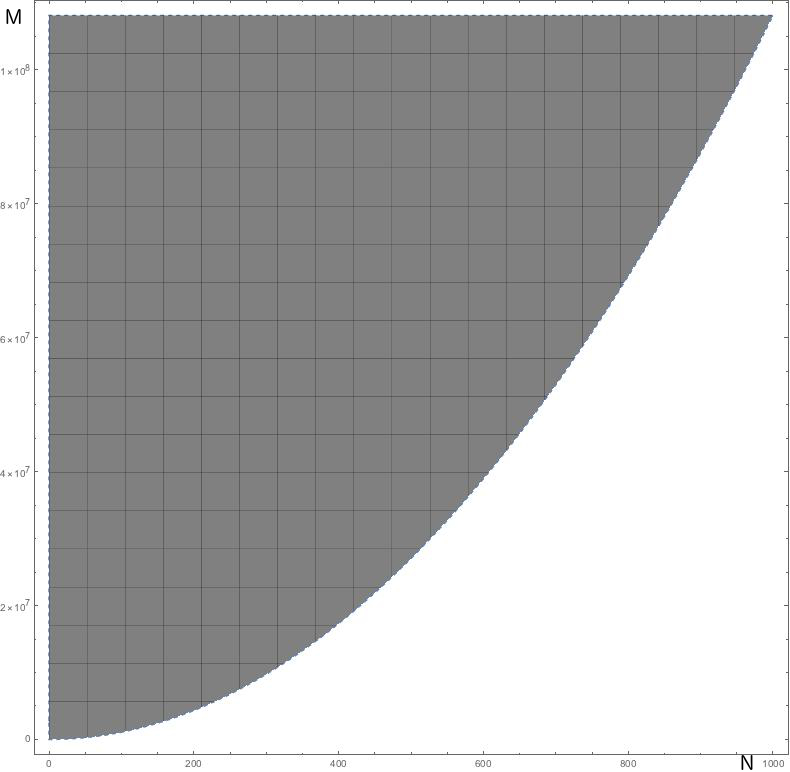}
    \caption{Shaded region on $N-M$ plane for which the theory has a critical point}
    \label{fig:reg}
\end{figure}

The Region on $N-M$ plane for which the theory has a critical point is in shaded region of Fig. \ref{fig:reg}

This critical exponents can also be calculated also in fixed dimension($d=3$) in the large $M$ limit. Where the coupling constants are $b\sim O(1/M),y\sim O(1/\sqrt{M}),g\sim O(1/\sqrt{M})$. This method is similar to what is described by Ma\cite{ma}. From this calculation we get for $M$- complex fields in fundamental representation of $SU(N)$,
\begin{align}
    \eta &=-\frac{1}{N M}\left[ 2.0264+2.1615(N^2-1) \right]\\
    \nu &=1-\frac{4.86}{N M}-\frac{4.32}{N M}(N^2-1)
\end{align}

This result matches with the already known results for $N=1$.\cite{HaLuMa,sakhi}

\section{Discussion and Conclusion}
From this analysis we found that for $SU(M)$ anti-ferromagnets there is a temperature driven phase transition for a very large $M$ compared to $N$(representation of the spin). This critical value $M_\text{crit}$ can be calculated for as a function of $N$.

The critical exponents of this second order phase transition are calculated in both $\epsilon$ expansion and in large-$M$ expansion. The next question one should ask is what are the phases that lie on the either side of the phase transition. 

It has already been discovered numerically that for $M=1$ there is no electro-weak phase transition at all for large value of $b/g^2$.\cite{laine,laine2,laine3} For large $M$ there is a phase transition. This phase transition corresponds to the breaking of the left over symmetry($U(M)$ flavour symmetry).\cite{fradkin,arnold2} The question still remains that what will be order parameter in that limit. It is known that those phases are connected to conventional Higgs and confinement phase.\cite{fradkin} One needs to study the lattice model rather than the coarse-grained theory to identify the phases.

All this analysis has been done when there is no topological term. The critical exponents can also be calculated if there is a topological term. The $U(1)$ case has been calculated recently\cite{sakhi} but $SU(N)\times U(1)$ case is not known. That can provide a better understanding of the topological phases in the actual lattice model for $M>1$ which is not known though the $M=1$ case has been studied.\cite{sachdev}

I plan to study in future the effect of the topological term in the Lagrangian and also the phases in the case of $M>1$ case on the lattice. There is also the case of non-bipartite lattice one may consider to study.

\section{Acknowledgement}
I thank Dr. Ganpathy Murthy for his help and valuable discussion to understand a number of shuttle issues about this problem. I am very thankful to Dr. Ribhu Kaul for introducing me to this area of Physics. I am very grateful to Dr. Michael Eides for his very helpful comments and discussions. I regard Dr. Peter Arnold for very helpful discussions.

I thank NSF(DMR-1306897) and University of Kentucky for supporting this project.

\newpage
\appendix
\section{Calculation of beta function in dimensional regularization}
We have to calculate the correction to the boson field propagator.
\begin{figure}[h]
    \includegraphics[scale=0.4]{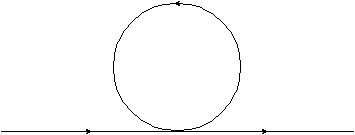}
    \includegraphics[scale=0.3]{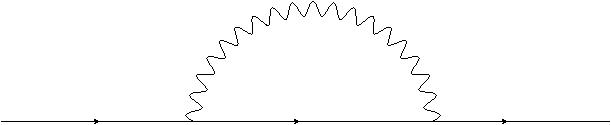}
    \includegraphics[scale=0.35]{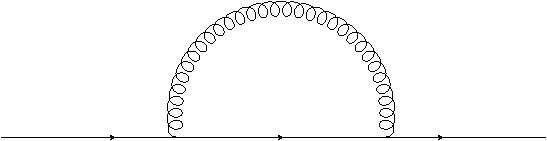}
    \includegraphics[scale=0.3]{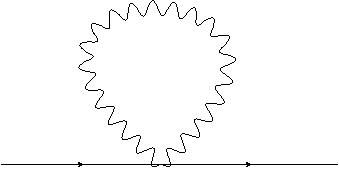}
    \includegraphics[scale=0.3]{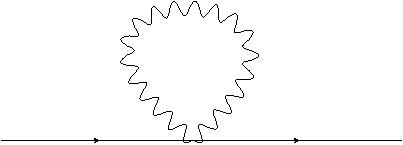}
    \caption{This diagrams (a1,a2,a3,a4,a5) contributes to the correction to the bosonic field propagator upto 1-loop order.}
\end{figure}

The above diagrams we need to calculate to find out the correction to the boson propagator. Where the propagator definitions are,
\begin{figure}[h]
    \includegraphics[scale=0.4]{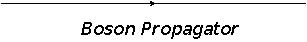}
    \includegraphics[scale=0.4]{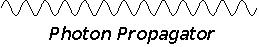}
    \includegraphics[scale=0.4]{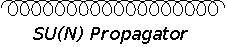}
    \includegraphics[scale=0.4]{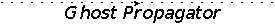}
    \caption{Propagators of the theory}
\end{figure}

One can calculate these diagrams easily to get in order $1/\epsilon$,
\begin{multline}
    diagram~a1\\
    =b\int\frac{d^dl}{(2\pi)^d}\frac{\delta_{kl}}{l^2+a}(\delta_{ik}\delta_{jl}+\delta_{il}\delta_{jk})\\
    =b(N+1)delta_{ij}\int\frac{d^d l}{(2\pi)^d}\frac{1}{l^2+a}\\
    =\frac{b(N+1)\delta_{ij}}{(4\pi)^{2-\epsilon/2}}\frac{\Gamma(1-d/2)}{\Gamma(1)}\left(\frac{1}{a}\right)^{1-d/2}\\
    =-\frac{ab(N+1)\delta_{ij}}{8\pi^2}\frac{1}{\epsilon}+O(\epsilon)
\end{multline}
\begin{multline}
    diagram~a2\\
    =-y^2\int\frac{d^d l}{(2\pi)^d}\frac{\delta_{kl}\left(\delta^{\mu\nu}-l^\mu l^\nu/l^2\right)(l+2k)^\nu (l+2k)^\mu \delta_{ik}\delta_{jl}}{l^2\left[ (l+k)^2+a \right]}\\
    =-4y^2\delta_{ij}\int\frac{d^d l}{(2\pi)^d}\frac{l^2k^2-(l\cdot  k)^2}{l^4\left[(l+k)^2+a\right]}\\
    =-4y^2\delta_{ij}\int dF_3\int\frac{d^d l}{(2\pi)^d}\frac{N}{\left[ x_1l^2+x_2 l^2+x_3(l+k)^2+x_3 a) \right]^3}\\
    =-4y^2\delta_{ij}(1-\frac{1}{d})k^2\int dF_3\int\frac{d^d q}{(2\pi)^d}\frac{q^2k^2}{\left[ q^2+\Delta \right]}\\
    =-\frac{3y^2\delta_{ij}}{8\pi^2}\frac{k^2}{\epsilon}+O(\epsilon)\\
\end{multline}
\begin{multline}
    diagram~a3\\
    =-g^2\int\frac{d^d l}{(2\pi)^d}\frac{\delta_{kl}\delta^{ab}(\delta^{\mu\nu}-\frac{l^\mu l^\nu}{l^2})(l+2k)_\mu(l+2k)_\nu T^a_{ik}T^b_{lj}}{l^2\left[ (l+k)^2+a \right]}\\
    =-\frac{g^2C_2(N)\delta_{ij}}{2}\int dF_3 \int\frac{d^d q}{(2\pi)^d}\frac{N}{(q^2+\Delta)^3}\\
\end{multline}
The $diagram~a4,diagram~a5$ can be calculated very simply and as the gauge theory is massless thus diagrams contribute zero. The results of these diagrams are,
\begin{align}
   diagram~a1&=-\frac{a b (N+1) \delta_{i j} }{8 \pi^2 \epsilon}+O(\epsilon)\\
   diagram~a2&=-\frac{3 y^2 \delta_{i j}k^2}{8\pi^2 \epsilon}+O(\epsilon)\\
   diagram~a3&=-\frac{3g^2 C_2(N)\delta_{i j}}{8\pi^2}k^2+O(\epsilon)\\
   diagram~a4&=0+O(\epsilon)\\
   diagram~a5&=0++O(\epsilon)
\end{align}

From this we can easily calculate the $Z$ values in the normalization as\cite{sred},
\begin{align}
Z_\psi&=1-\frac{3 y^2}{8\pi^2 \epsilon}-\frac{3g^2 C_2(N)}{8 \pi^2 \epsilon}\\
Z_a&=1-\frac{b(N+1)}{8\pi^2 \epsilon}
\end{align}

The diagrams that will contribute to the $U(1)$ gauge propagator,
\begin{figure}
    \includegraphics[scale=0.35]{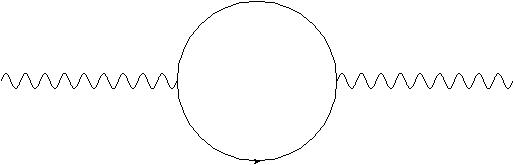}
    \includegraphics[scale=0.4]{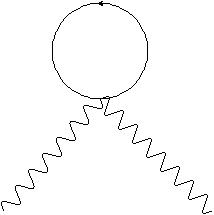}
    \caption{This diagrams (b1,b2) contributes to the correction to the $U(1)$ field propagator upto 1-loop order.}
\end{figure}
We can calculate the diagram to be,
\begin{multline}
    diagram~b1\\
    =-y^2\int\frac{d^d l}{(2\pi)^d}\frac{\delta_{ik}\delta_{jk}(l+2k)^\mu(k+2l)^\nu\delta_{ij}\delta_{kl}}{(l^2+a)\left[ (l+k)^2+a \right]}\\
    =-y^2 N\int dF_2\int\frac{d^d l}{(2\pi)^d)}\frac{N^{\mu\nu}}{\left[ x_1(l^2+a)+x_2(l+k)^2+x_2 a \right]}\\
    =-y^2N\int dF_2\int\frac{d^d q}{(2\pi)^d}\frac{\frac{4}{d}\delta^{\mu\nu}q^2+(2x_2-1)^2k^\mu k^\nu}{\left[ q^2+\Delta \right]^2}\\
    =\frac{y^2 N}{8\pi^2 \epsilon}\int_0^1dx\left[ 2\delta^{\mu\nu}(x(1-x)k^2+a)-(2x-1)^2k^\mu k^\nu \right]+O(\epsilon)\\
    =\frac{y^2N}{24\pi^2\epsilon}\left[ k^2\delta^{\mu\nu}-k^\mu k^\nu \right]+\frac{y^2 Na}{4\pi^2 \epsilon}\delta^{\mu\nu}+O(\epsilon)
\end{multline}

\begin{multline}
    diagram~b2=2y^2\int\frac{d^d l}{(2\pi)^d}\frac{\delta_{ij}delta_{ij\delta^{\mu\nu}}}{l^2+a}\\
    =-\frac{aNy^2\delta^{\mu\nu}}{4\pi^2 \epsilon}
\end{multline}

As we can see these diagram add to give a transverse field as expected and also,
\begin{align}
    Z_B=1+\frac{Ny^2}{24\pi^2 \epsilon}
\end{align}

Similarly one need to calculate the correction to the non-abelian gauge propagator.

\begin{figure}
    \includegraphics[scale=0.35]{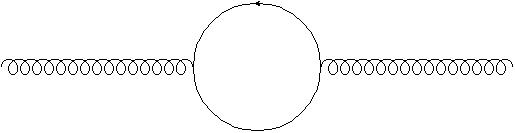}
    \includegraphics[scale=0.35]{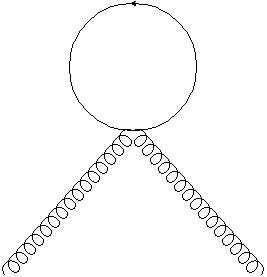}
    \includegraphics[scale=0.25]{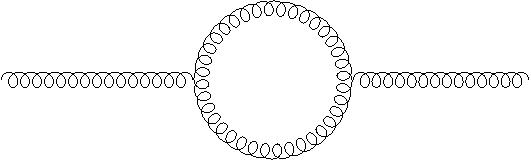}
    \includegraphics[scale=0.3]{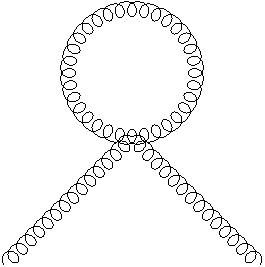}
    \includegraphics[scale=0.35]{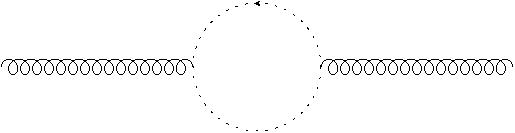}
    \caption{This diagrams (c1,c2,c3,c4,c5) contributes to the correction to the SU(N) field propagator upto 1-loop order.}
\end{figure}

We similarly can calculate in the Feynman gauge \cite{pes} to calculate these diagrams,
\begin{multline}
    diagram~c1\\
    =-g^2\int\frac{d^d l}{(2\pi)^d}\frac{\delta_{ij}\delta_{kl}(2l+k)^\mu(2l+k)^\nu T^a_{ik}T^b_{lj}}{(l^2+a)\left[ (l+k)^2+a \right]}\\
    =\frac{g^2\delta^{ab}}{48\pi^2\epsilon}\left[ k^2\delta^{\mu\nu}-k^\mu k^\nu \right]+\frac{g^2 a \delta^{\mu\nu}\delta^{ab}}{8\pi^2\epsilon}
\end{multline}
\begin{multline}
    diagram~c2=2g^2\delta^{\mu\nu}T^a_{ik}T^b_{kj}\int\frac{d^d l}{(2\pi)^d}\frac{\delta_{ij}}{l^2+a}\\
    =-\frac{ag^2\delta^{\mu\nu}\delta^{ab}}{8\pi^2\epsilon}
\end{multline}
The calculation for the $diagram~c3,diagram~c4,diagram~c5$ is straight forward in transverse gauge,\cite{pes}
\begin{align}
    diagram~(c3+c4+c5)=-\frac{13}{3}\frac{Ng^2\delta^{ab}}{16\pi^2 \epsilon}(\delta^{\mu\nu}k^2-k^\mu k^\nu)
\end{align}

This gives,
\begin{equation}
    Z_W=1-\frac{g^2}{16\pi^2\epsilon}\left[ \frac{13}{3}N-\frac{1}{3} \right]
\end{equation}

Next is the correction to the four boson vertex,

\begin{figure}
    \includegraphics[scale=0.28]{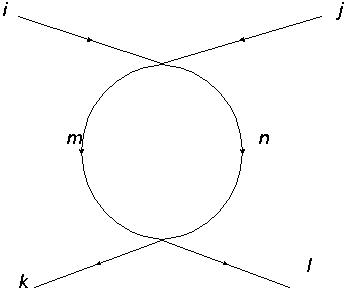}
    \includegraphics[scale=0.07]{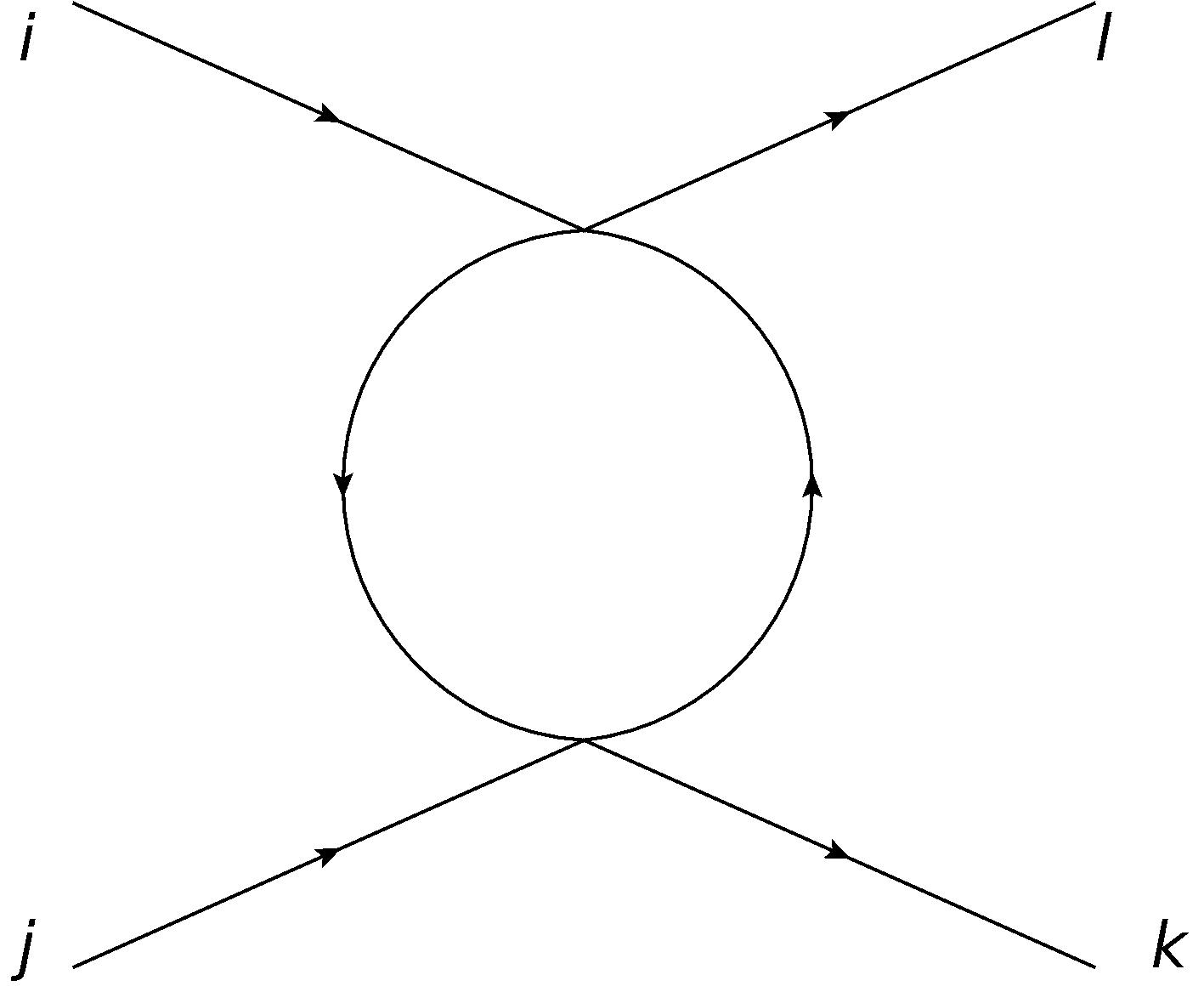}
    \includegraphics[scale=0.07]{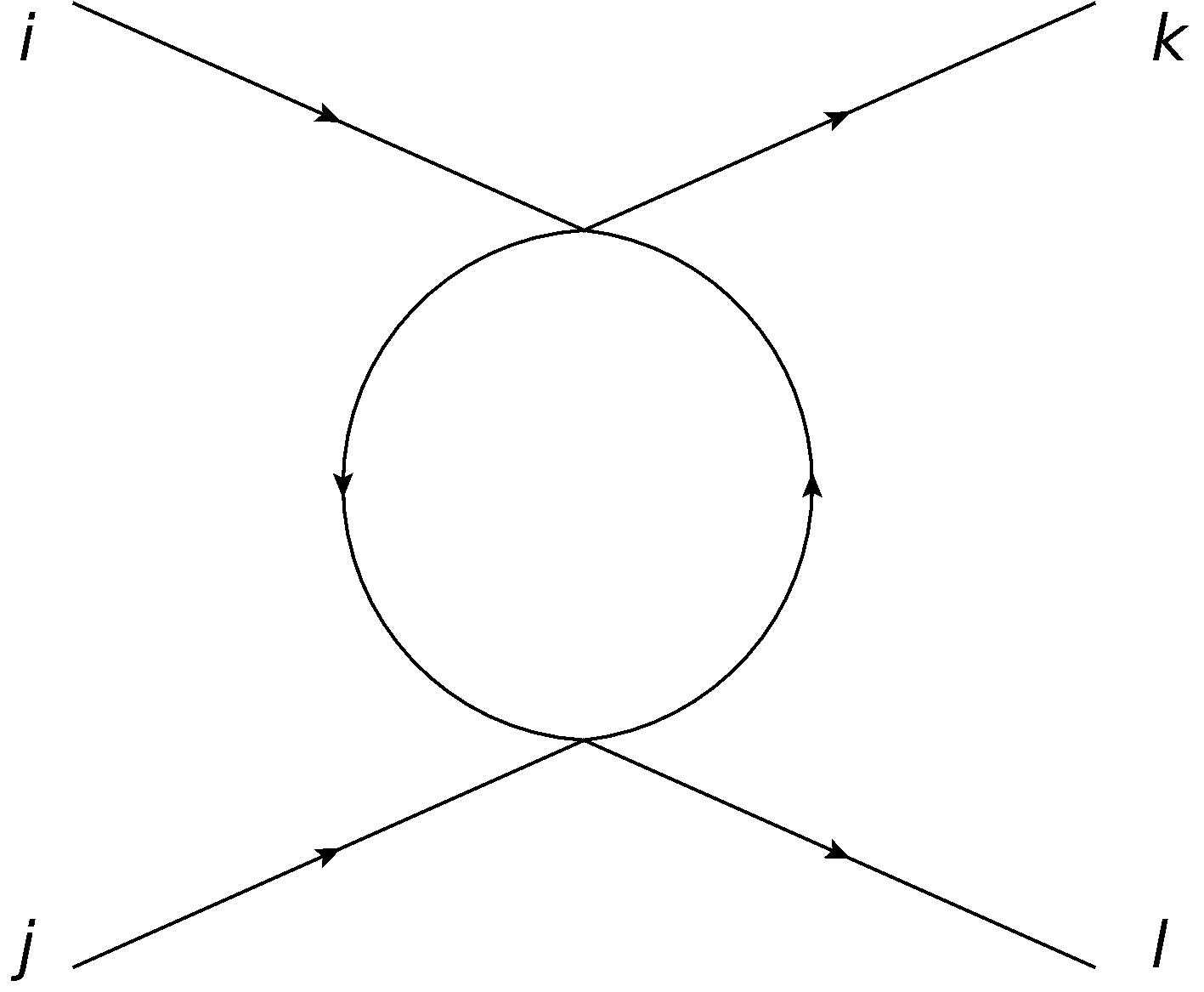}
    \includegraphics[scale=0.07]{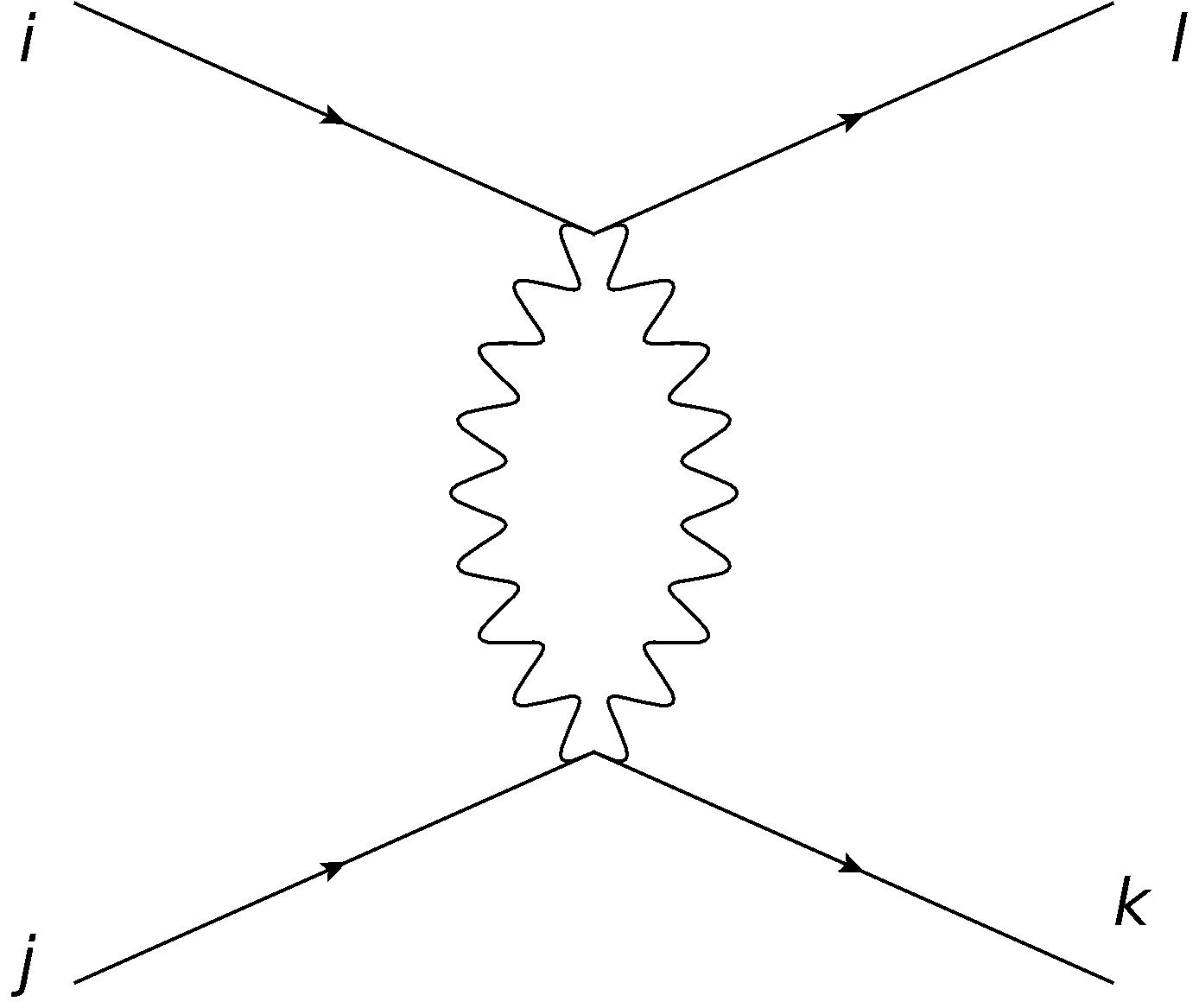}
    \includegraphics[scale=0.07]{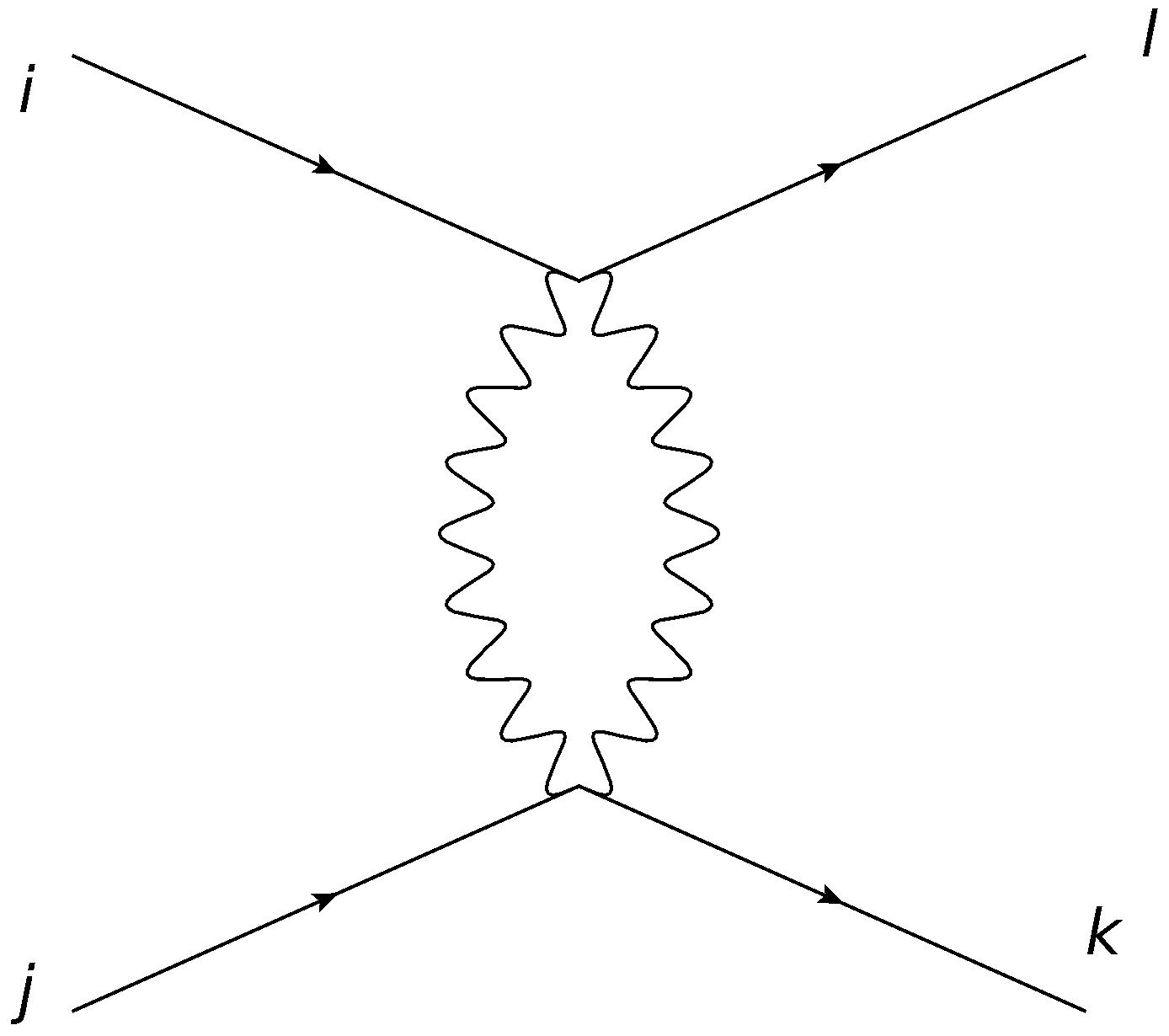}
    \includegraphics[scale=0.07]{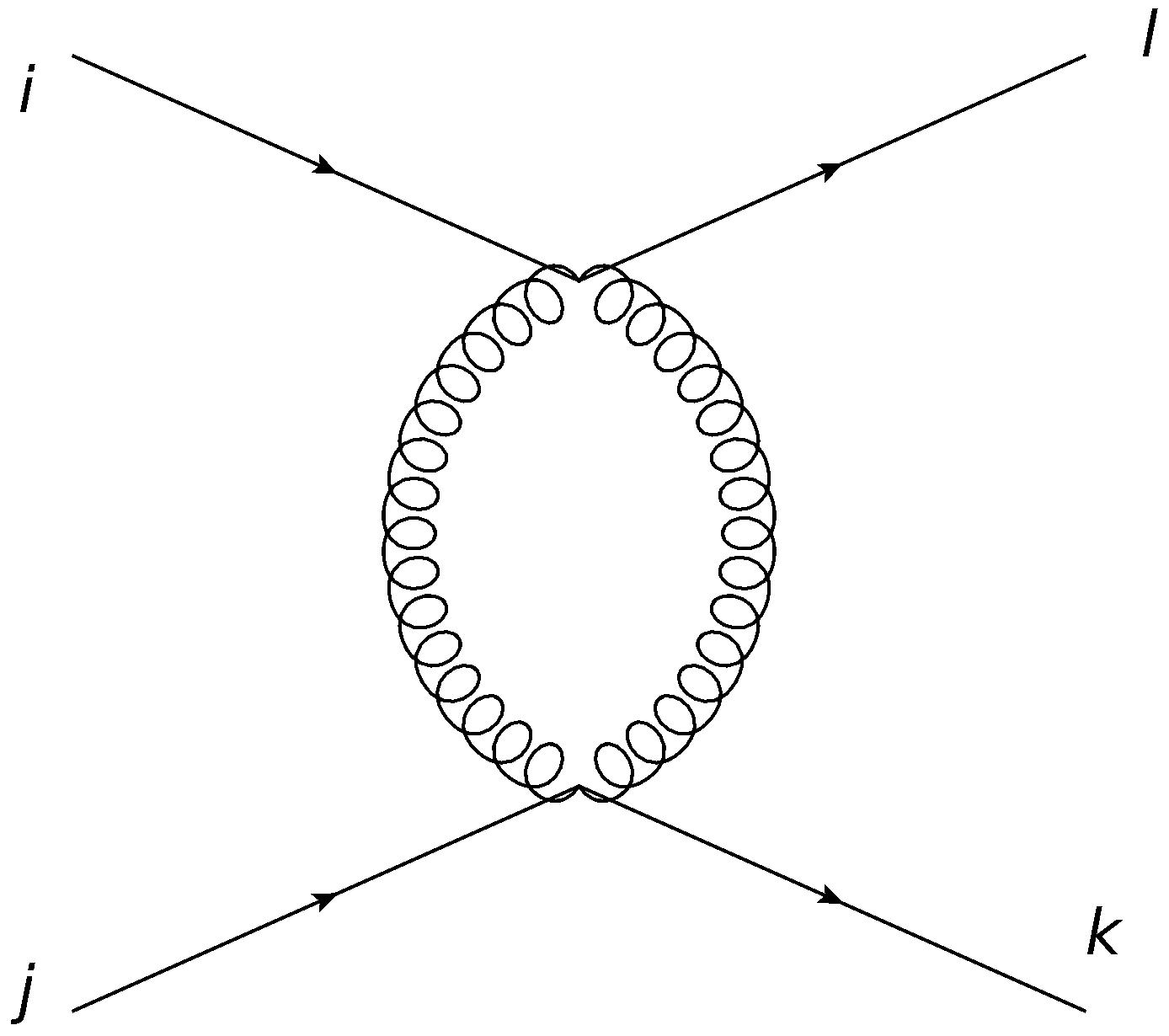}
    \includegraphics[scale=0.07]{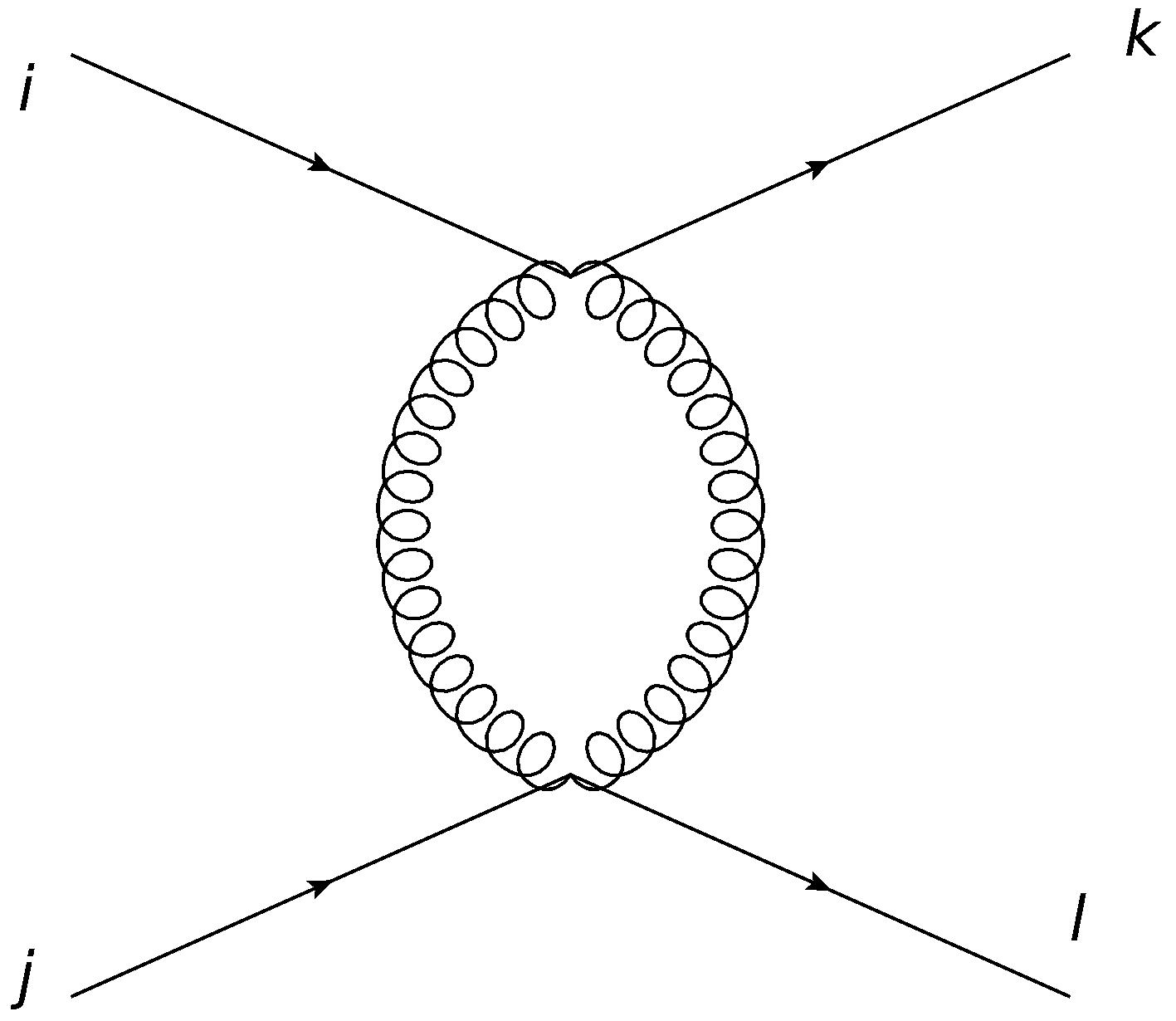}
    \includegraphics[scale=0.07]{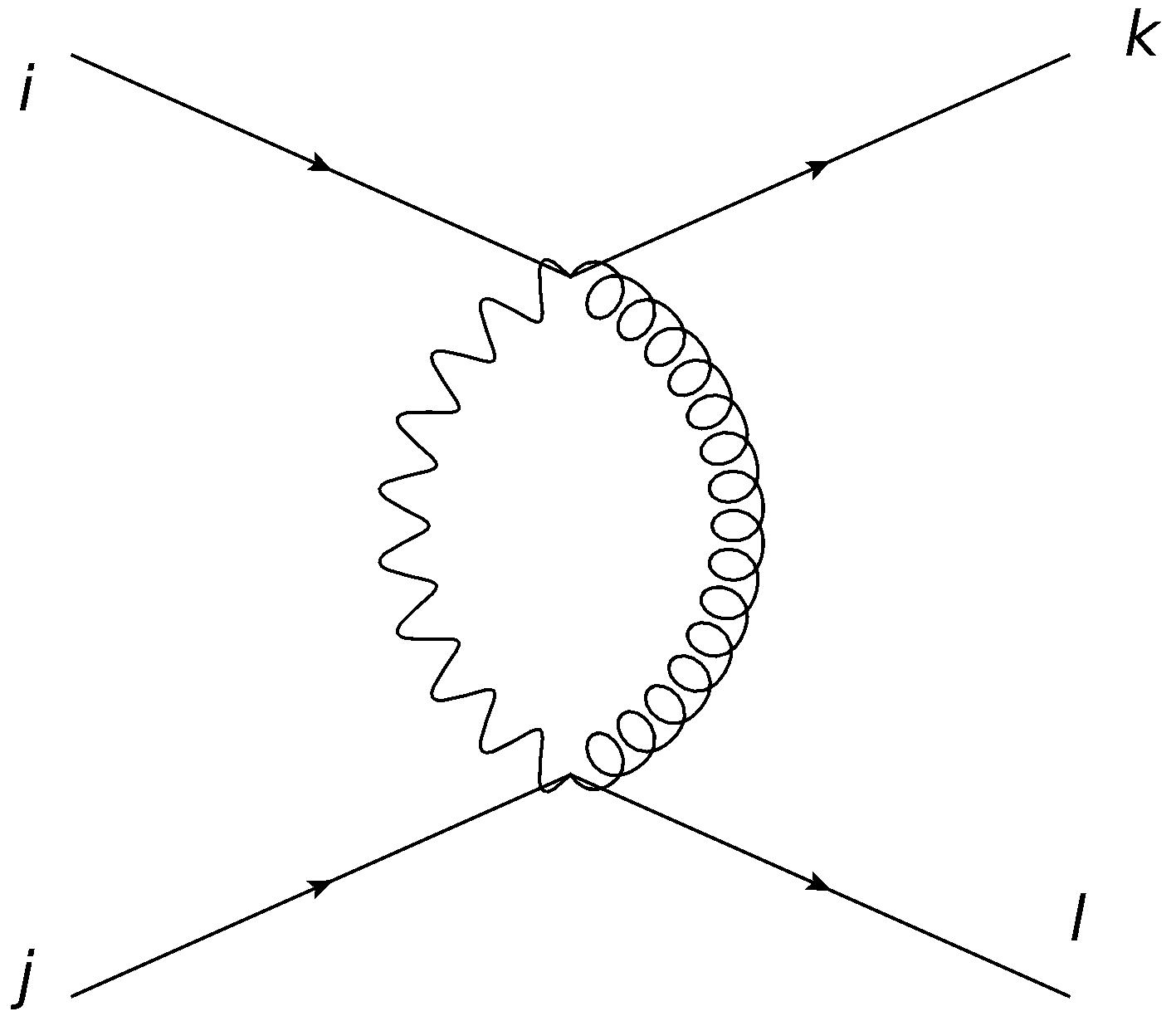}
    \includegraphics[scale=0.07]{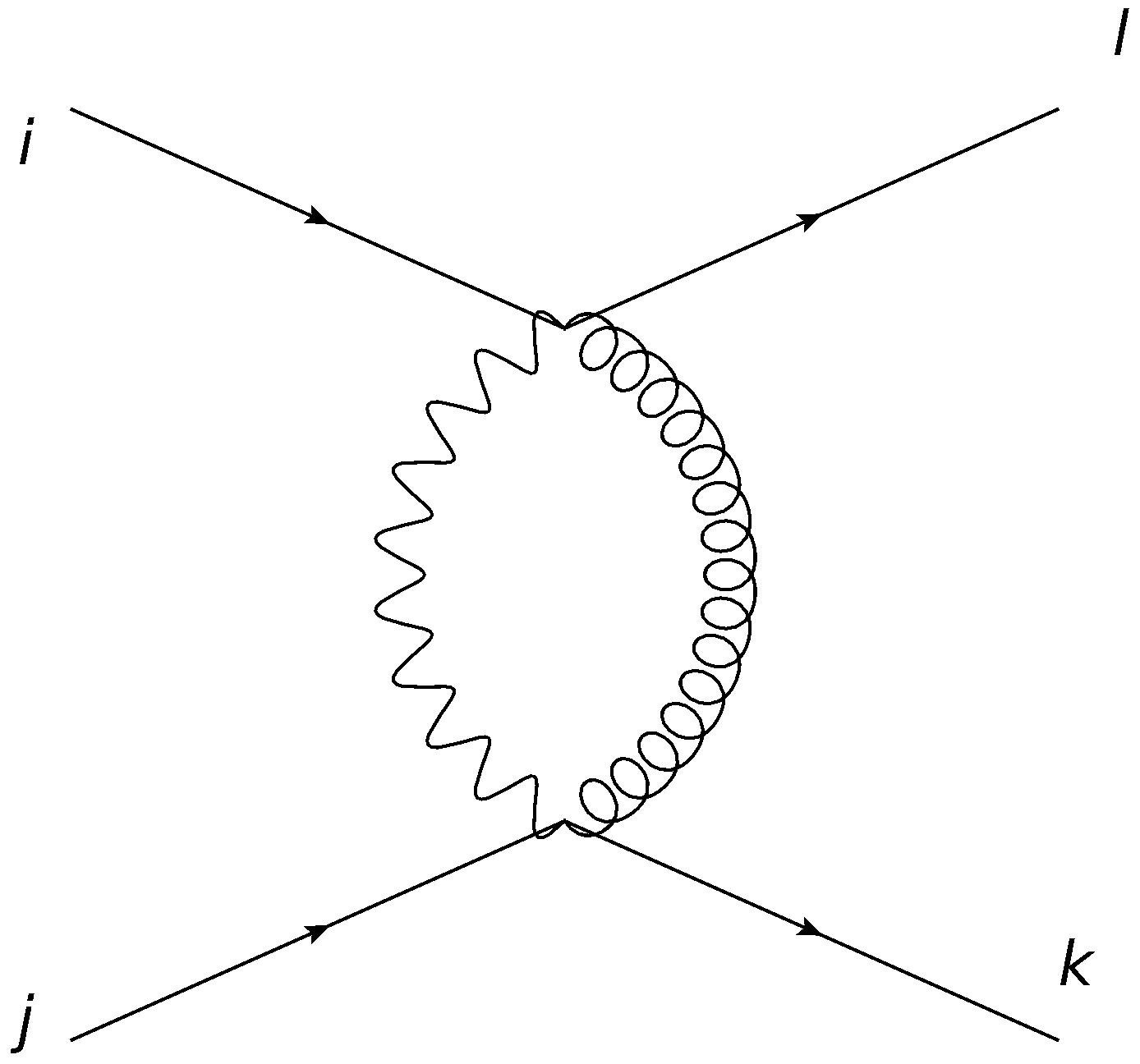}
    \caption{This diagrams (d1,d2,d3,d4,d5,d6,d7,d8,d9) contributes to the four boson vertex upto 1-loop order.}
\end{figure}

This diagrams can be calculated as,
\begin{multline}
    diagram~(d1+d2+d3)\\
    =-b^2(N+4)(\delta_{ik}\delta_{jl}+\delta_{il}\delta_{jk})\int\frac{d^d l}{(2\pi)^d}\frac{1}{(l^2+a)^2}\\
    =-\frac{b^2(N+4)}{8\pi^\epsilon}(\delta_{ik}\delta_{jl}+\delta_{il}\delta_{jk})+O(\epsilon)
\end{multline}

\begin{multline}
    diagram~(d4+d5)\\
    =-2y^2(\delta_{ik}\delta_{jl}+\delta_{il}\delta_{jk})\int\frac{d^d l}{(2\pi)^d}\frac{\left( \delta^{\mu\nu}-l^\mu l^\nu/l^2 \right)(\delta_{\mu\nu}-l_\mu l_\nu/l^2)}{l^4}\\
    =-\frac{3y^4}{4\pi^2 \epsilon}(\delta_{ik}\delta_{jl}+\delta_{il}\delta_{jk})+O(\epsilon)
\end{multline}

\begin{multline}
    diagram~(d6+d7)=-\frac{3g^4}{32\pi^2\epsilon}\left(\frac{N^3+N^2-4N+2}{N^2}\right)(\delta_{ik}\delta_{jl}+\delta_{il}\delta_{jk})
\end{multline}

\begin{multline}
    diagram~(d8+d9)\\
    =-2(gy)^2\left(\frac{N-1}{N}\right)(d-1)(\delta_{ik}\delta_{jl}+\delta_{il}\delta_{jk})\int \frac{d^d l}{(2\pi)^d}\frac{1}{l^4}\\
    =-\frac{3g^2y^2}{4\pi^2\epsilon}\left(\frac{N-1}{N}\right)(\delta_{ik}\delta_{jl}+\delta_{il}\delta_{jk})
\end{multline}

\begin{equation}
Z_b b=b-\frac{b^2(N+4)}{8\pi^2 \epsilon}-\frac{3y^4}{4\pi^2\epsilon}-\frac{3g^4}{32\pi^2\epsilon}\left( \frac{N^3+N^2-4N+2}{N^2} \right)    
\end{equation}

Next we need to look for 3-point $U(1)$-boson-boson vertex,
\begin{figure}
    \includegraphics[scale=0.1]{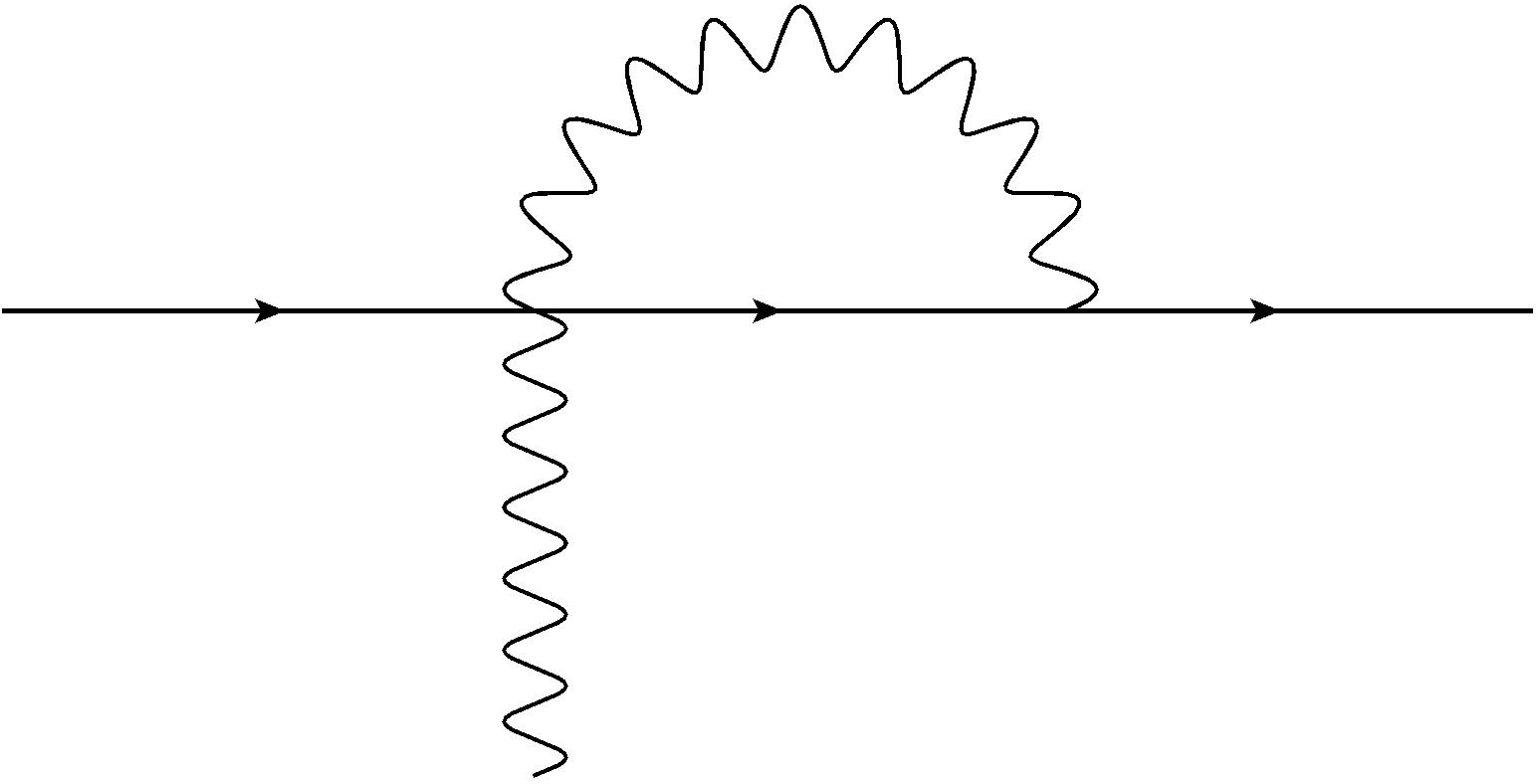}
    \includegraphics[scale=0.08]{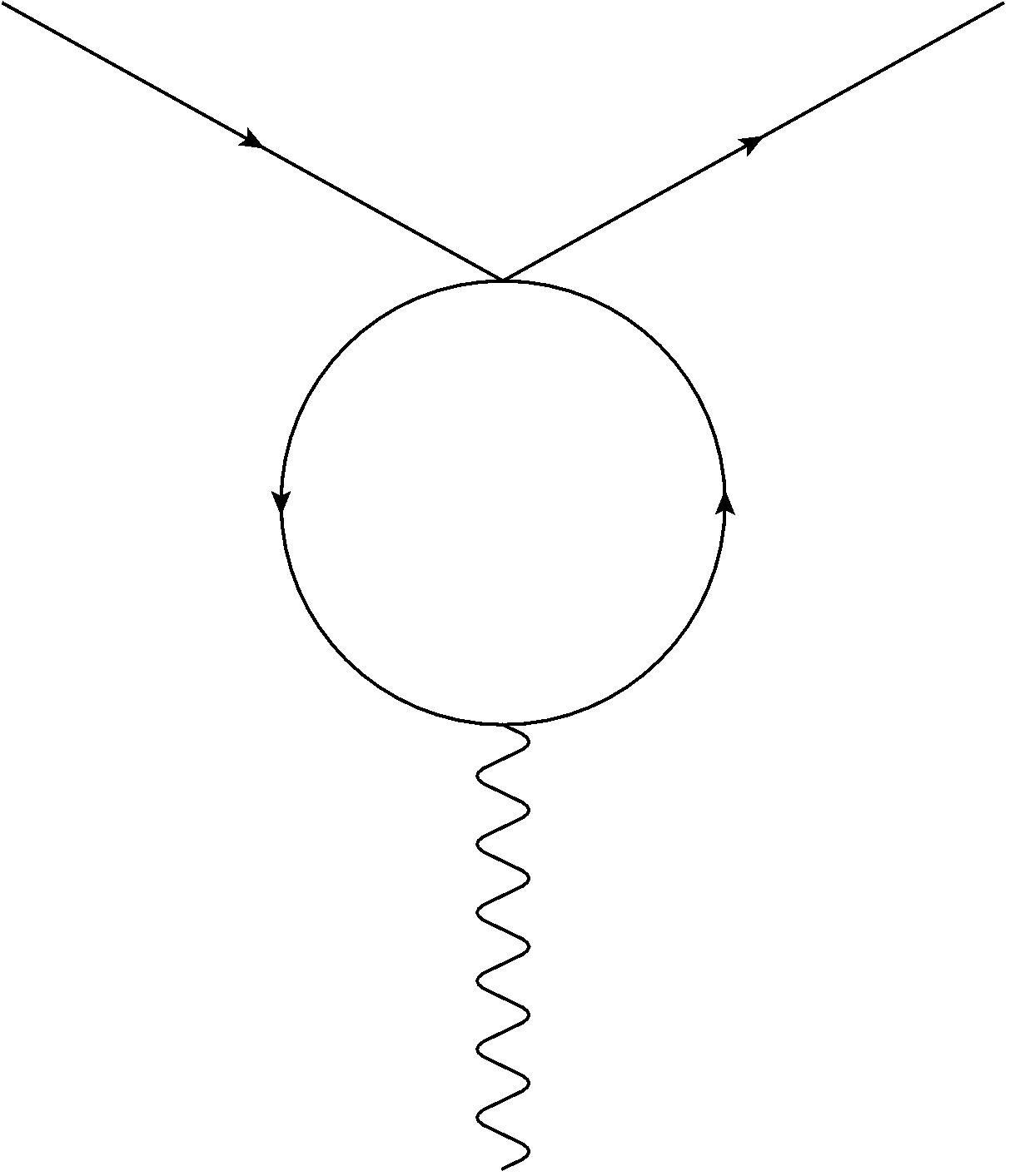}
    \includegraphics[scale=0.08]{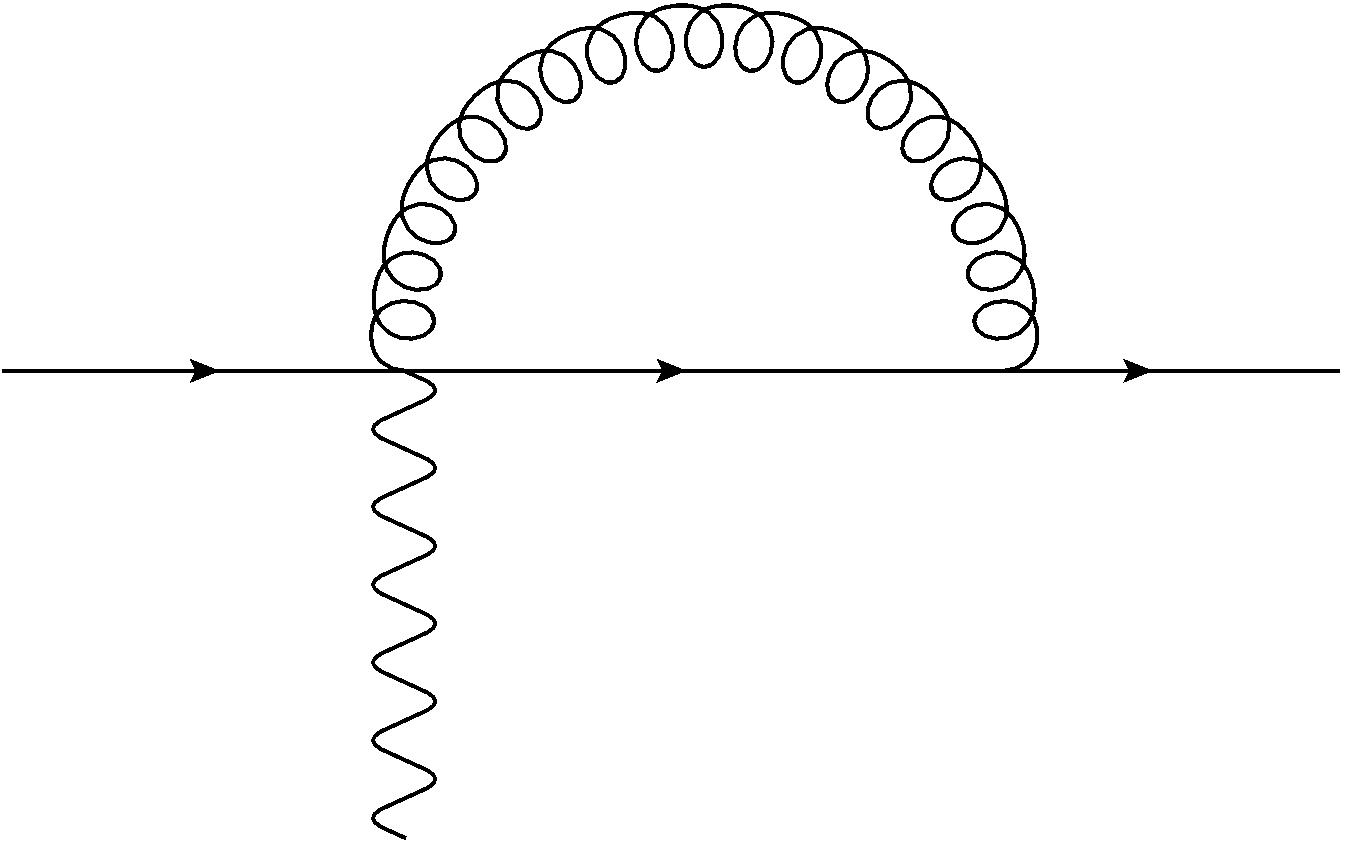}
    \caption{This diagrams (e1,e2,e3) contributes to the correction to the $U(1)$-boson-boson vertex upto 1-loop order.}
\end{figure}
The value of the diagrams will be,
\begin{multline}
    diagram~e1=4y^3\delta_{ij}(1-1/d)p^\rho\int\frac{d^d l}{(2\pi)^d}\frac{1}{l^2\left[ (l+p)^2+a \right]}\\
    =\frac{3y^3\delta_{ij}p^\rho}{8\pi^2\epsilon}+O(\epsilon)
\end{multline}

\begin{multline}
    diagram~e2\\
    =\frac{by(N+1)\delta_{ij}}{2}\int\frac{d^d l}{(2\pi)^d}\frac{(2l+p)^\mu}{(l^2+a)\left[ (l+p)^2+a \right]}\\
    =\frac{by(N+1)\delta_{ij}}{2}\int_0^1(1-2x)dx\int\frac{d^d l}{(2\pi)^d}\frac{1}{\left[ q^2+\Delta \right]^2}=0
\end{multline}

\begin{multline}
    diagram~e3=\\
    g^2 y\int\frac{d^d l}{(2\pi)^d}\frac{(\delta^{\mu\nu}-\frac{l^\mu l^\nu}{l^2})(l+2p)_\nu \delta_{m l}}{l^2l^2\left[ (l+p)^2+a \right]}\delta^{n m}T^a_{i m}T^a_{l j}\\
    =-\frac{3g^2y C_2(N)}{8\pi^2 \epsilon}p^\mu \delta_{ij}+O(\epsilon)
\end{multline}
Thus, we can write down $Z_1$ as,
\begin{equation}
    Z_1=1-\frac{3y^2}{8\pi^2\epsilon}-\frac{3g^2y}{8 \pi^2 \epsilon}\left(\frac{N^2-1}{2N} \right)
\end{equation}

Next we need to look for 3-point $SU(N)$-boson-boson vertex,

\begin{figure}
    \includegraphics[scale=0.075]{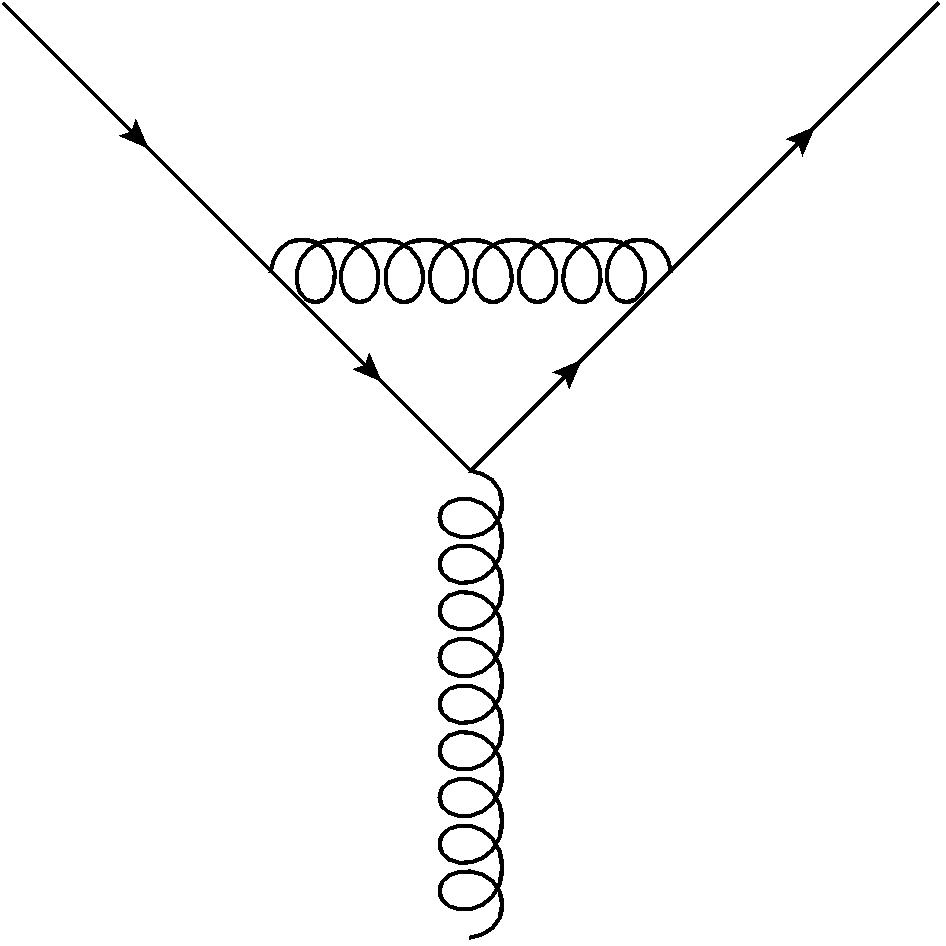}
    \includegraphics[scale=0.09]{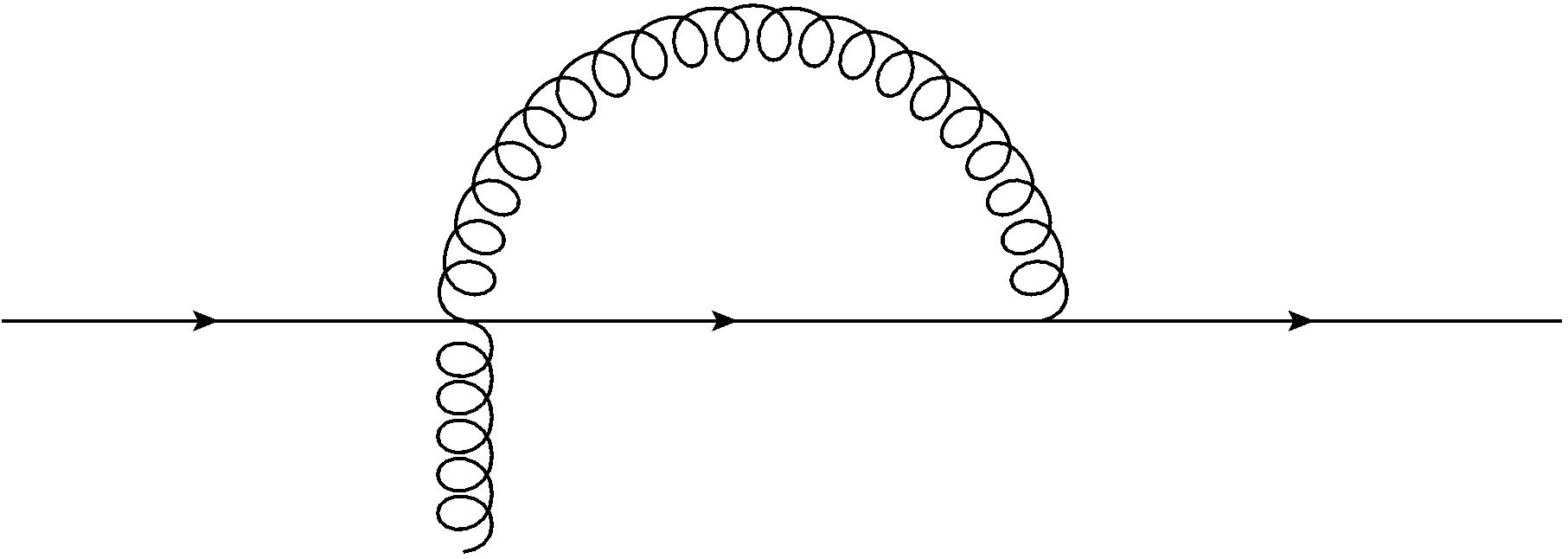}
    \includegraphics[scale=0.08]{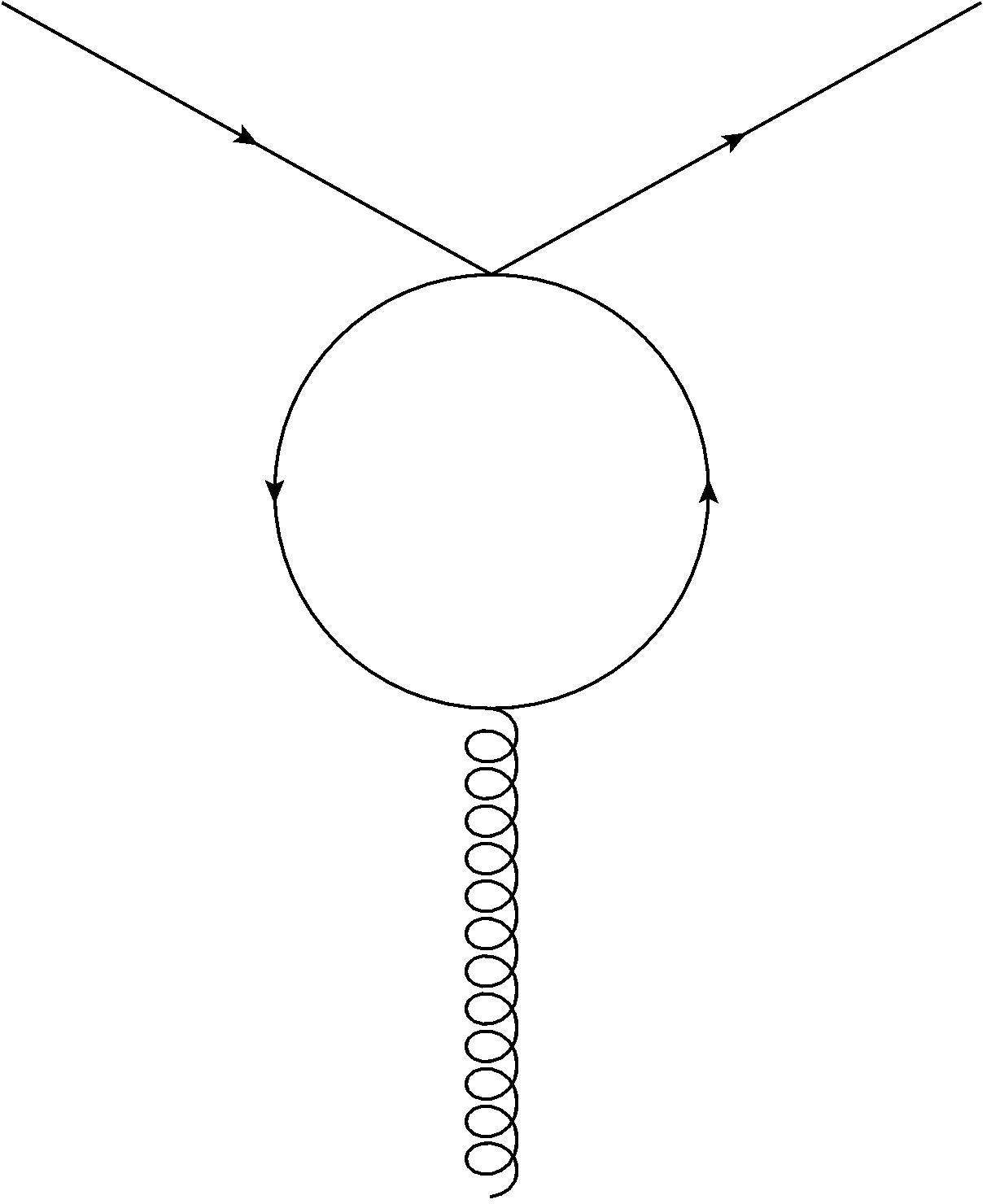}
    \includegraphics[scale=0.08]{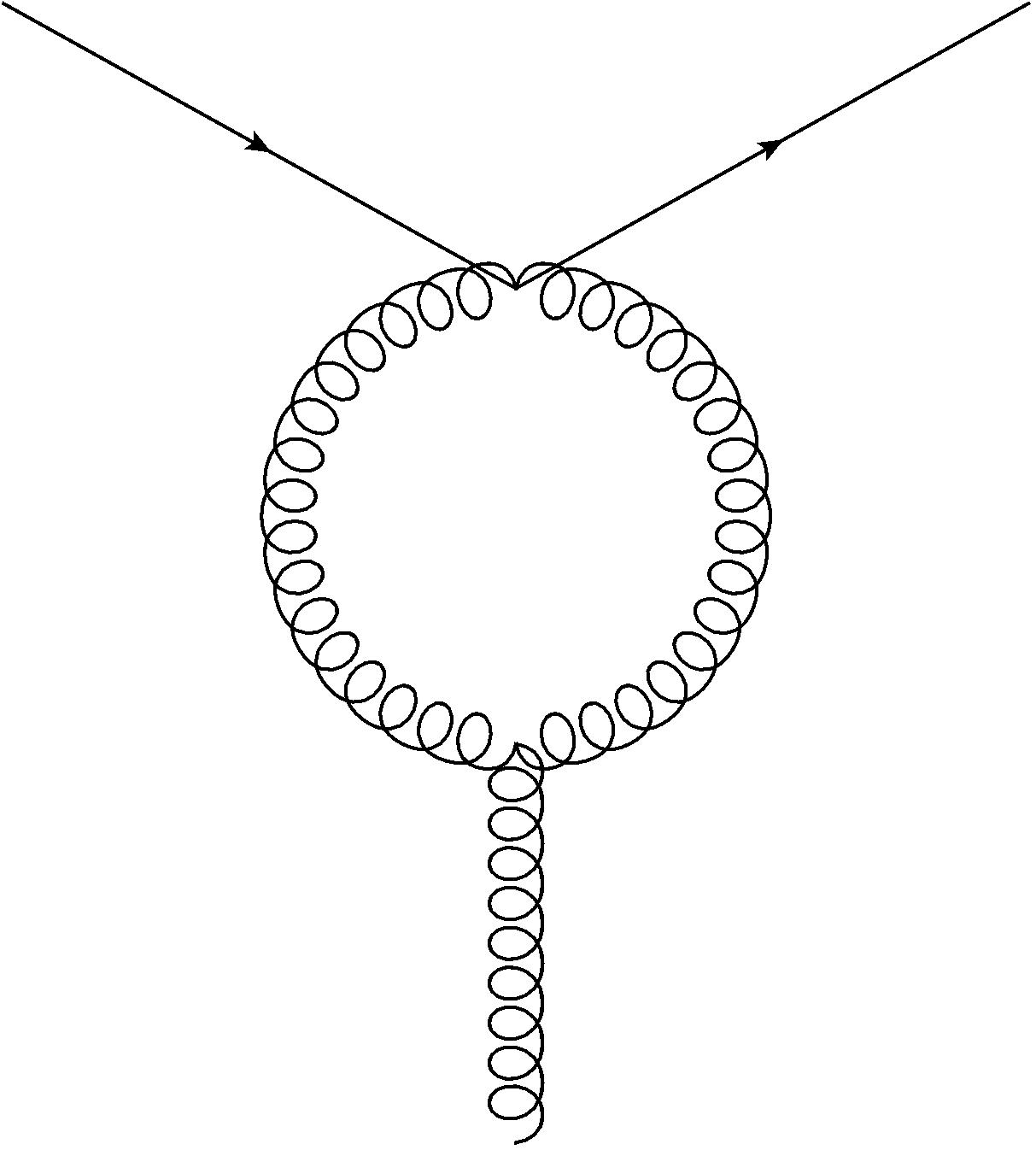}
    \includegraphics[scale=0.06]{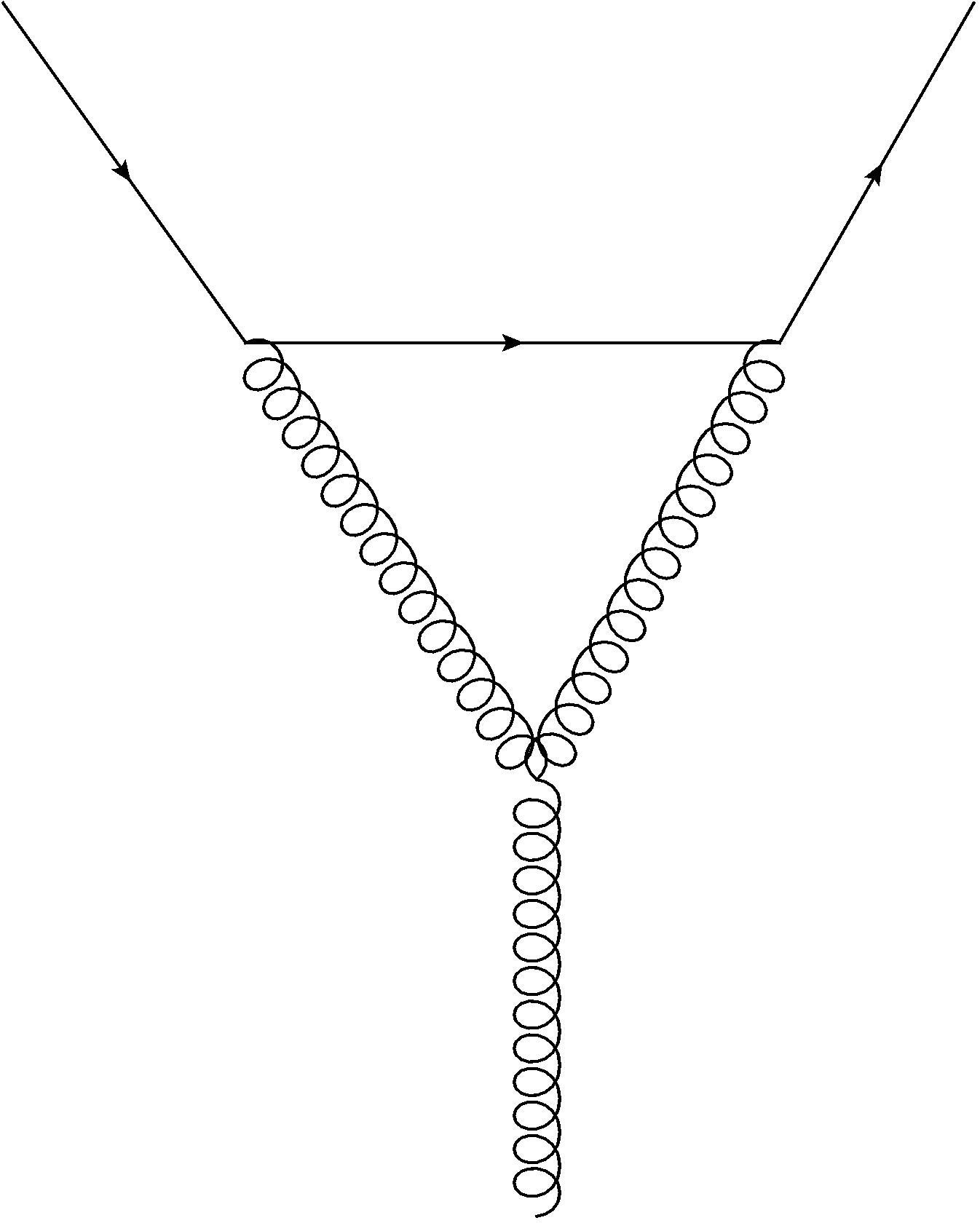}
    \includegraphics[scale=0.1]{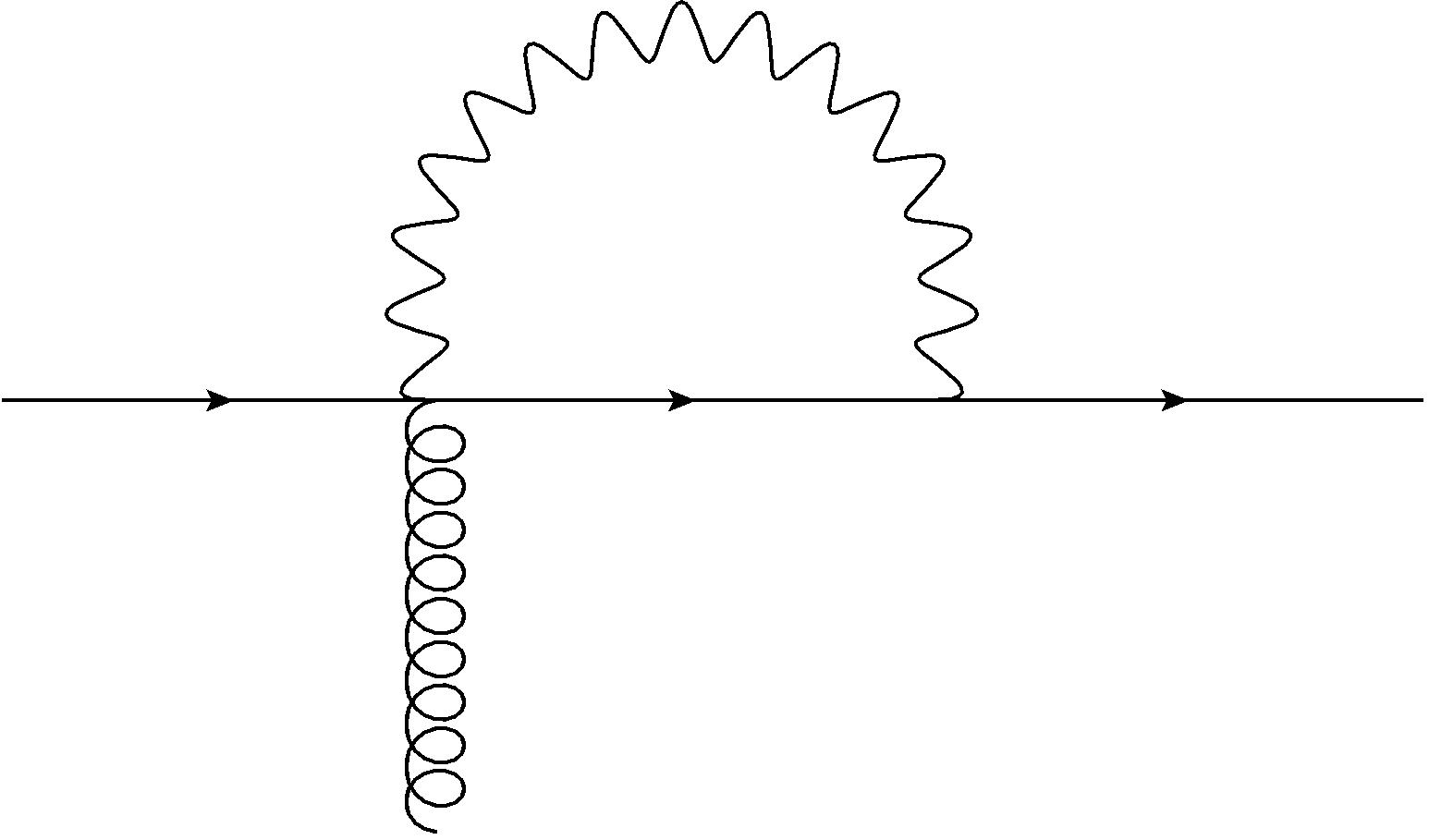}
    \caption{This diagrams (f1,f2,f3,f4) contributes to the correction to the $SU(N)$-boson-boson upto 1-loop order.}
\end{figure}

The evaluation of this diagrams will be,
\begin{align}
    diagram~f1=0\text{~~~by gauge}
\end{align}
\begin{equation}
    diagram~f2=\frac{3g^3}{32\pi^2\epsilon}\left(\frac{N^2-2}{N}\right)T^a_{ij}p^\mu+O(\epsilon)
\end{equation}
$diagram~f3$ is zero similarly to the $U(1)$ case.
\begin{align}
    diagram~f4&=0\text{~~~Can be shown}
\end{align}
Similarly $diagram~f5$ is zero for transverse gauge.
\begin{multline}
    diagram~f6=\frac{4gy^2}{2}T^a_{ij}\int\frac{d^d l}{(2\pi)^d}\frac{l^2 p^\mu-(l\cdot p)l^\mu}{l^2 l^2\left[(l+p)^2+a\right]}\\
    =\frac{3 g y^2}{16 \pi \epsilon}T^a_{ij}p^\mu
\end{multline}
Now, as before we can calculate $Z_2$ here as,
\begin{equation}
    Z_2=1-\frac{3g^2}{32\pi^2N \epsilon}(N^2-2)-\frac{3y^2}{8\pi^2\epsilon}
\end{equation}
Using,
\begin{align}
    \psi_R&=Z_\psi^{1/2}\psi_B\\
    B_R&=Z_B^{1/2}B_B\\
    y_R&=Z_1Z_\psi^{-1}Z_B^{-1/2}y_B\\
    g_R&=Z_2Z_\psi^{-1}Z_W^{-1/2}g_B\\
    a_R&=Z_aZ_\psi^{-1}a\\
    b_R&=bZ_b Z_\psi^{-2}
\end{align}
One can calculate the beta functions.

\section{Critical exponents in large $M$ limit}
Next we need to calculate the critical exponents in the philosophy of \cite{ma}. I will also discuss the calculation done by ma but only in $d=3$.

We will calculate $\eta$ and $\gamma$ directly then from that using scaling relation one can calculate $\nu=\gamma/(2-\eta)$. For this lets start with when we have no gauge field. Then the action looks like,
\begin{equation}
    S=\int d^3x\left[ \abs{\vec{\nabla}\psi}+\frac{b}{2}\abs{\psi}^4 \right]
\end{equation}
Now, we can introduce a hubbard-stratonovich field $\chi$ to reduce the four point vertex into a a three point vertex, i.e.
\begin{multline}
    \exp\left[ -\frac{b}{2}\int d^3\abs{\psi}^4 \right]=\\
    \frac{1}{\sqrt{2\pi b}}\int \mathcal{D}\chi \exp\left[ -\int d^3x\exp\left\{ \frac{1}{2b}\chi^2+i\chi\abs{\psi} \right\} \right]
\end{multline}
\begin{figure}[h]
    \centering
    \includegraphics[scale=0.15]{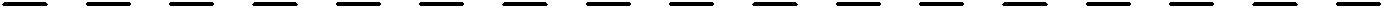}
    \caption{Propagator of the Hubbard-Stratonovich field}
\end{figure}
Thus the propagator of the $\chi$ field is $b$. Now we want to calculate the correction to the $\psi$ propagator to $O(1/M)$.
\begin{figure}[h]
    \includegraphics[scale=0.09]{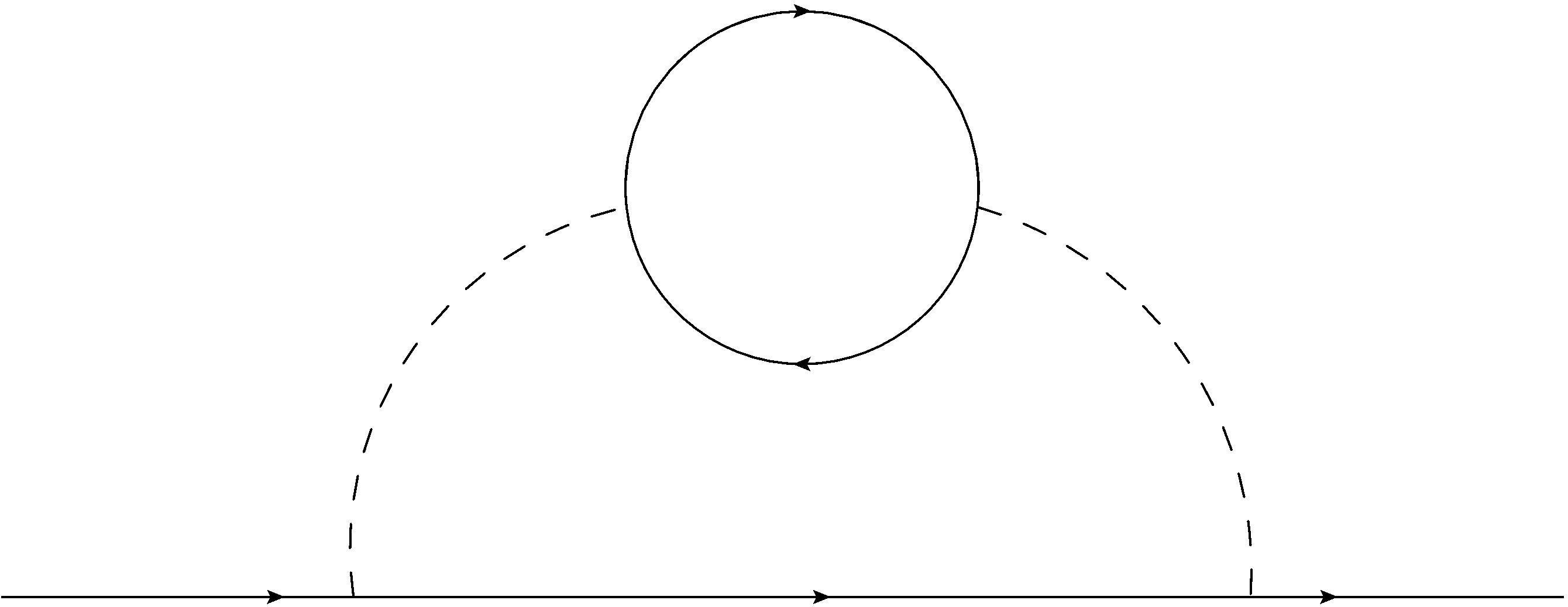}
    \includegraphics[scale=0.09]{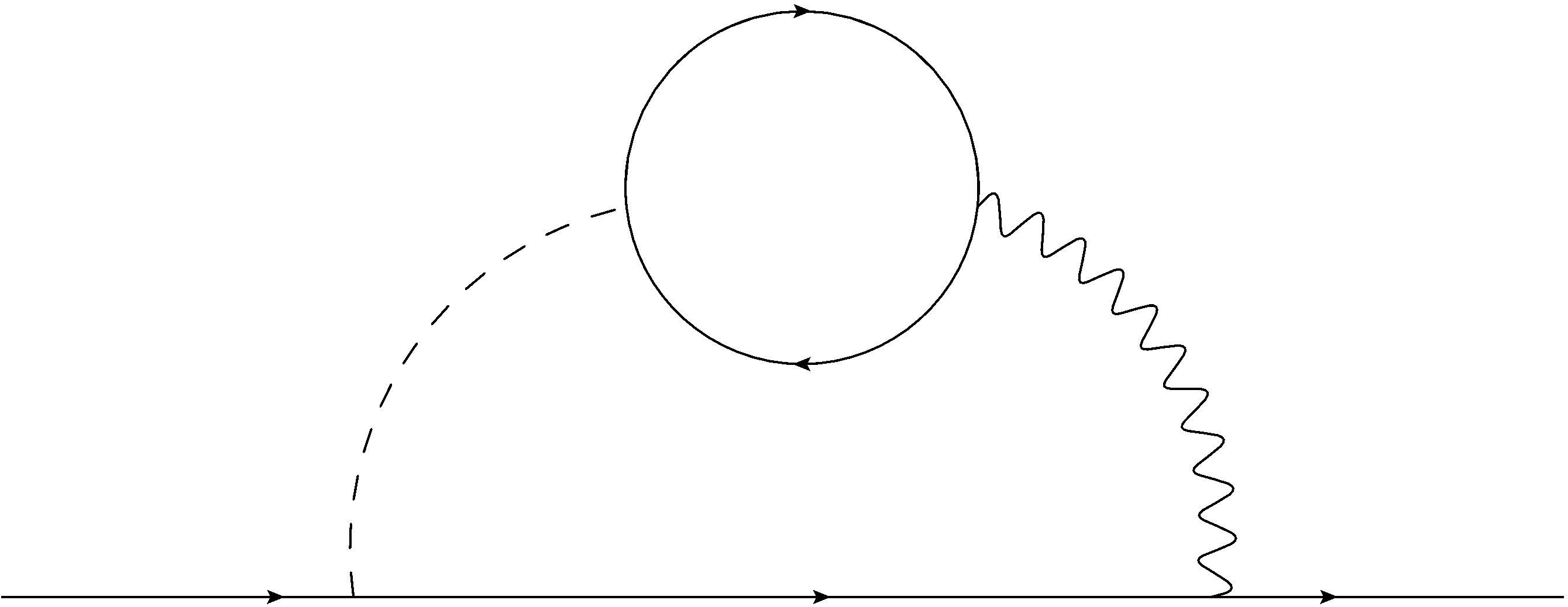}
    \caption{Sample diagrams g1,g2 contributing to the $\psi$ propagator}
\end{figure}
We can add as many boson loop as we need in the hubbard propagator. Thus, adding all those,
\begin{equation}
    b+b(-NM\pi)+b(-NM\pi)^2+b(-NM\pi)^3+...=b(1+NMb\pi)
\end{equation}
Where, $\pi$ is called the fundamental bubble,
\begin{equation}
    \pi(p)=\int \frac{d^3 l}{(2\pi)^3}\frac{1}{l^2(l+p)^2}=\frac{\Gamma(1/2)}{(4\pi)^{3/2}p}\beta(1/2,1/2)
\end{equation}
Thus the correction to the propagator is,
\begin{equation}
    \Sigma(k)=\int\frac{d^3 l}{(2\pi)^3}\frac{-b}{1+NMb\pi(l)}\frac{1}{(l+k)^2}
\end{equation}
\subsection{Critical exponent $\eta$}
Now, the critical exponent $\eta$ is defined as(in mass zero limit, the case here),
\begin{equation}
    G^{-1}=k^2+\Sigma(k)-\Sigma(0)\sim k^{2-\eta}
\end{equation}
Thus, we need to calculate
\begin{equation}
    \Sigma(k)-\Sigma(0)=\int\frac{d^3 l}{(2\pi)^3}\frac{b}{1+NMb\pi(l)}\left(\frac{1}{l^2}-\frac{1}{(l+k)^2}\right)
\end{equation}
From this we need to find the coefficient of $-k^2\log(k)$, which can be done easily as only small $l$ region that contributes and $\pi(l)$ diverges for small $l$ and we can neglect $1$ w.r.t. $\pi(l)$ and found to be
\begin{equation}
    \eta=\frac{4}{3\pi^2 N M}
\end{equation}
Next if we introduce a $U(1)$ gauge field then we can see the following diagrams contribute to $O(1/M)$,
\begin{figure}[h]
    \includegraphics[scale=0.09]{g3.jpg}
    \includegraphics[scale=0.11]{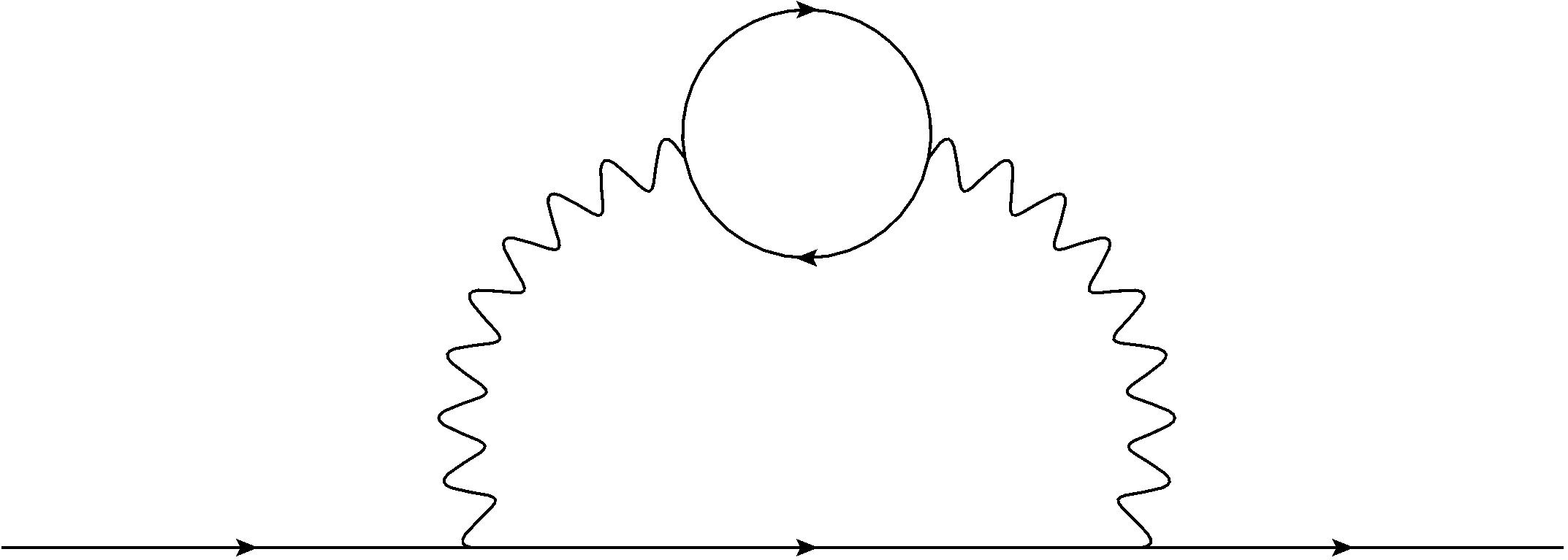}
    \caption{Sample diagrams g3,g4 contributing to $O(1/M)$}
\end{figure}
Lets concentrate on the mixed diagram,
\begin{figure}[h]
    \centering
    \includegraphics[scale=0.09]{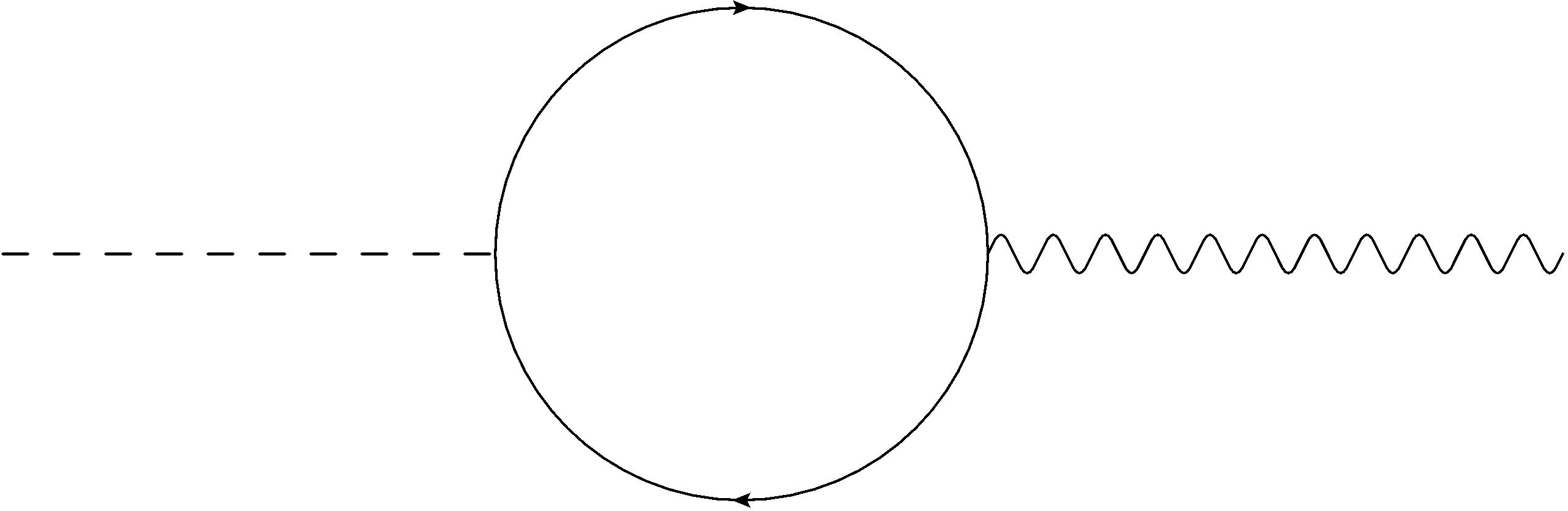}
    \caption{Mixed contribution diagram~g5}
\end{figure}
This mixed term goes to zero very simply. Thus we just need to calculate the correction to the $U(1)$ propagator,
\begin{figure}
    \includegraphics[scale=0.09]{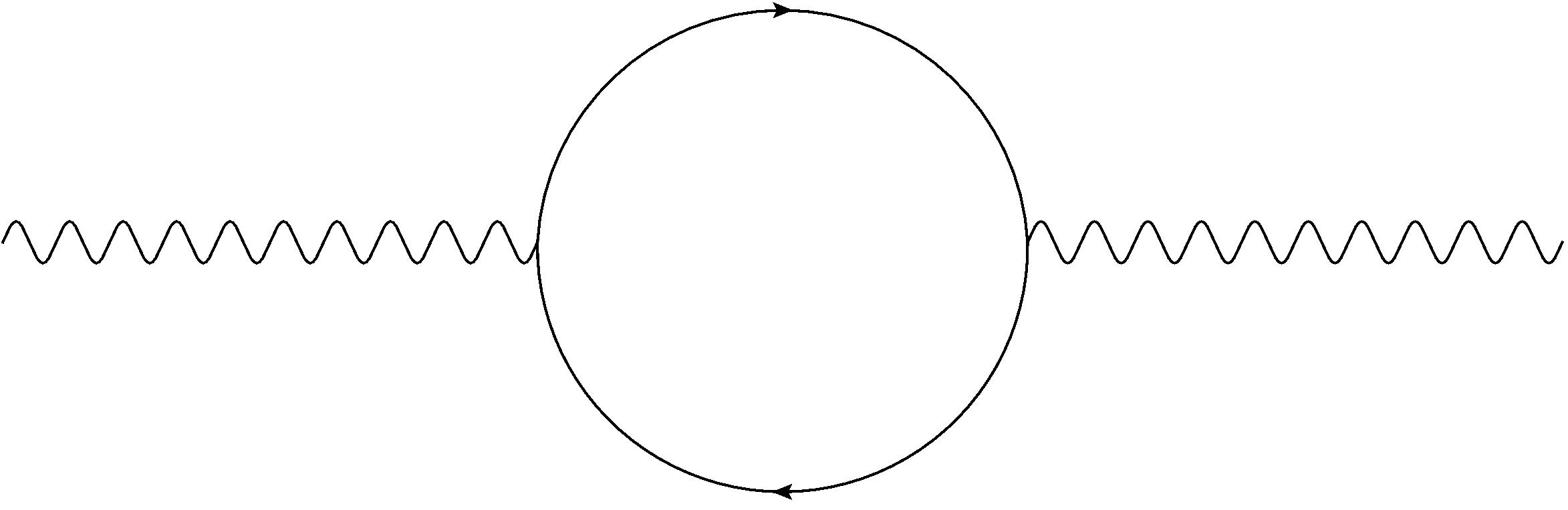}
    \includegraphics[scale=0.09]{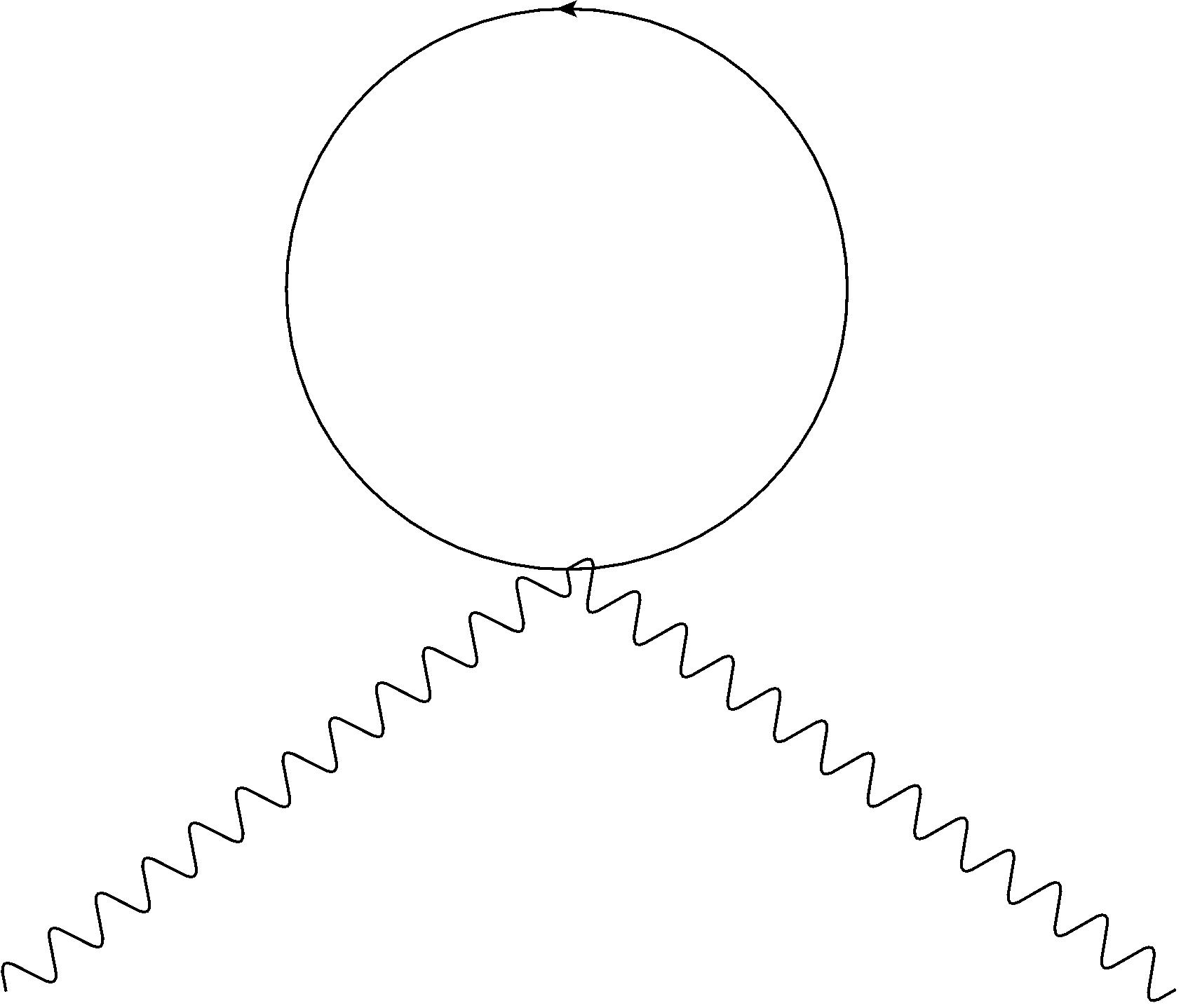}
    \caption{Basic bubble for the correction to the $U(1)$ Propagator diagram g6,g7}
\end{figure}
\begin{multline}
    diagram~g6+g7=\\
    N My^2\int\frac{d^3}{(2\pi)^3}\left[\frac{(p+2l)_\mu(p+2l)_\nu-2\delta_{\mu\nu}}{l^2((l+p)^2)}\right]\\
    =-N M y^2\frac{\Gamma(1/2)}{(4\pi)^{3/2}p}\frac{\sqrt{\pi}\Gamma(1/2)}{2\Gamma(2)}\left[ p^2\delta_{\mu\nu}-p_\mu p_\nu \right]\\
    \text{(This we get using the Feymann trick)}\\
    =N M y^2 \pi(p)\left[ p^2\delta_{\mu\nu}-p_\mu p_\nu \right]
\end{multline}
Thus the corrected $U(1)$ Propagator will be,
\begin{equation}
    \frac{1}{q^2(1-NMy^2\pi(p))}\left[ \delta_{\mu\nu}-\frac{q_\mu q_\nu}{q^2} \right]
\end{equation}
And we will represent the as double line diagram,
\begin{figure}
    \centering
    \includegraphics[scale=0.09]{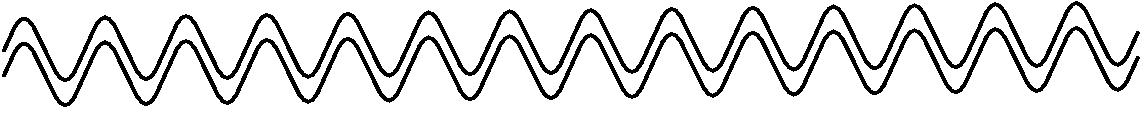}
    \caption{Corrected $U(1)$ propagator upto $O(1/M)$}
\end{figure}
Now, we want to collect the correction to the $\psi$ propagator upto $O(1/M)$,
\begin{figure}
    \includegraphics[scale=0.09]{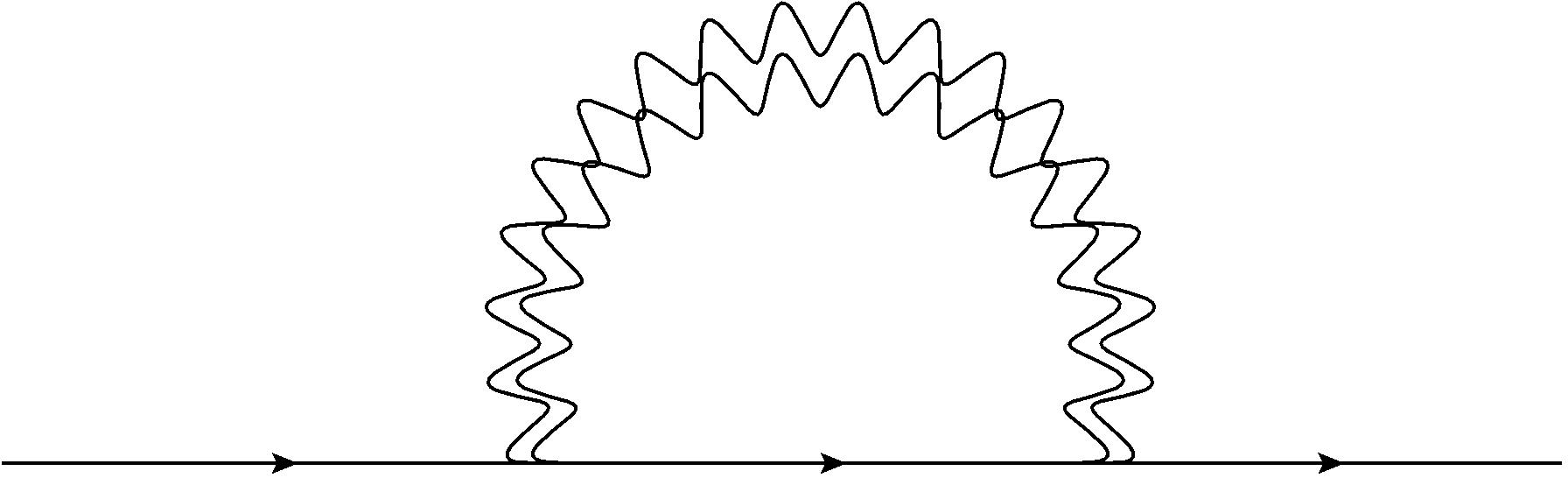}
    \includegraphics[scale=0.09]{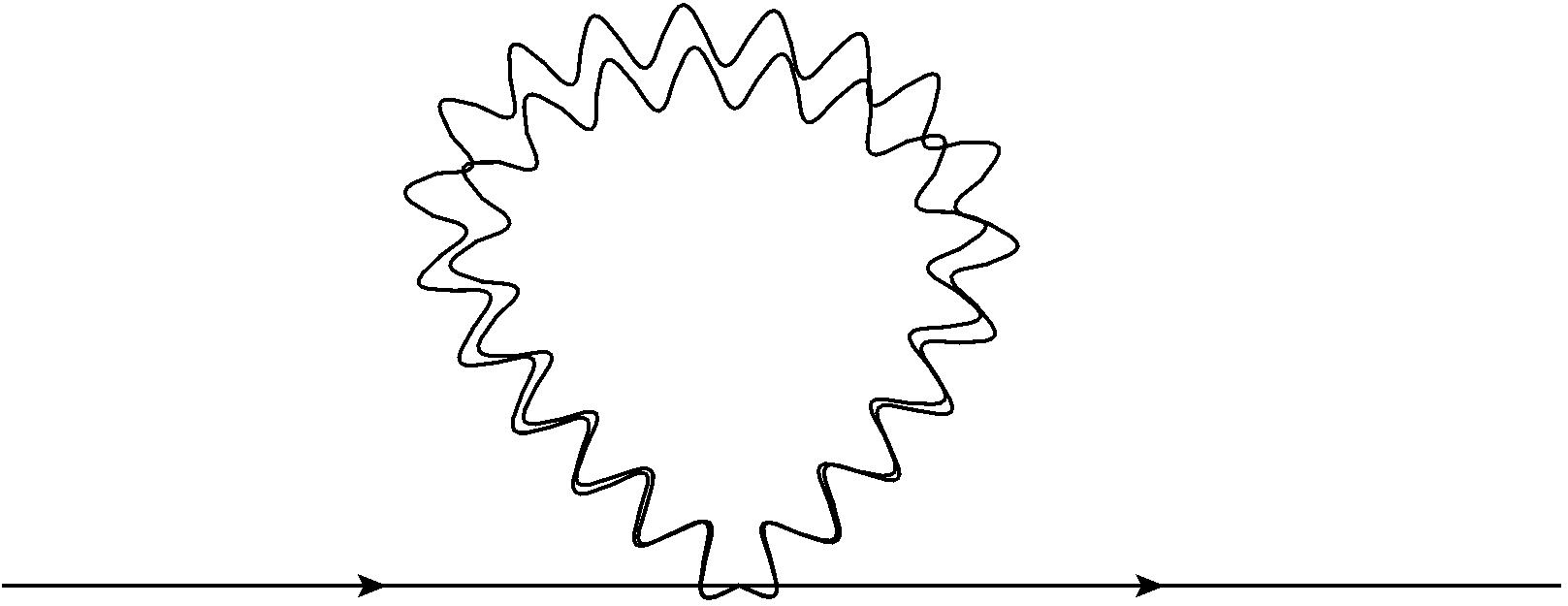}
    \caption{$U(1)$ contribution to the propagator correction upto $O(1/M)$ diagram g8,g9}
\end{figure}
\begin{multline}
    diagram~g8+g9\\
    =\int\frac{d^3l}{(2\pi)^3}\frac{y^2 \delta^{\alpha\beta}}{1-NM y^2\pi(l)}\left[ \frac{4(l^2k^2-(l\cdot k)^2)-4l^2(l+k)^2}{l^4(l+k)^2} \right]
\end{multline}
Again we find the $-k^2\log k$ coefficient,
\begin{equation}
    \eta_B=-\frac{16}{3}\frac{1}{2\pi NM}\frac{1}{2\pi\Gamma(1)}\frac{2\Gamma(2)(4\pi)^{3/2}}{\sqrt{\pi}\Gamma(1/2)\Gamma(1/2)}=-\frac{2.16152}{N M}
\end{equation}
Next we need to introduce $SU(N)$ gauge field. Again the mixed diagrams will cancel. Now the correction to the $SU(N)$ propagator will be,
\begin{figure}
    \includegraphics[scale=0.085]{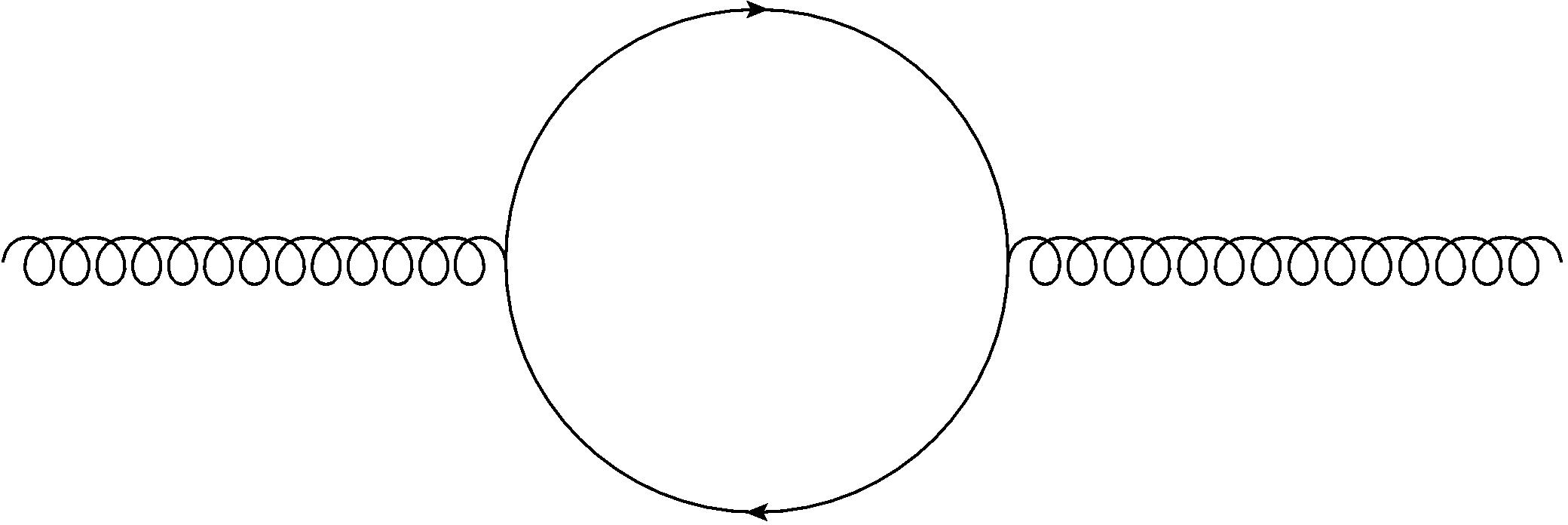}
    \includegraphics[scale=0.085]{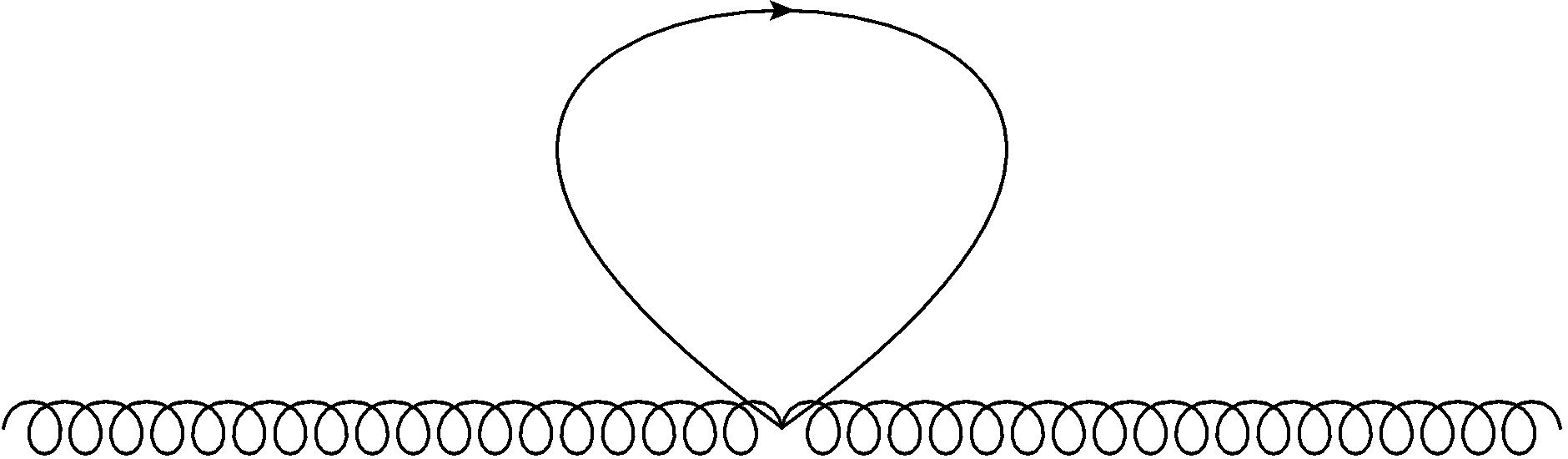}
    \includegraphics[scale=0.085]{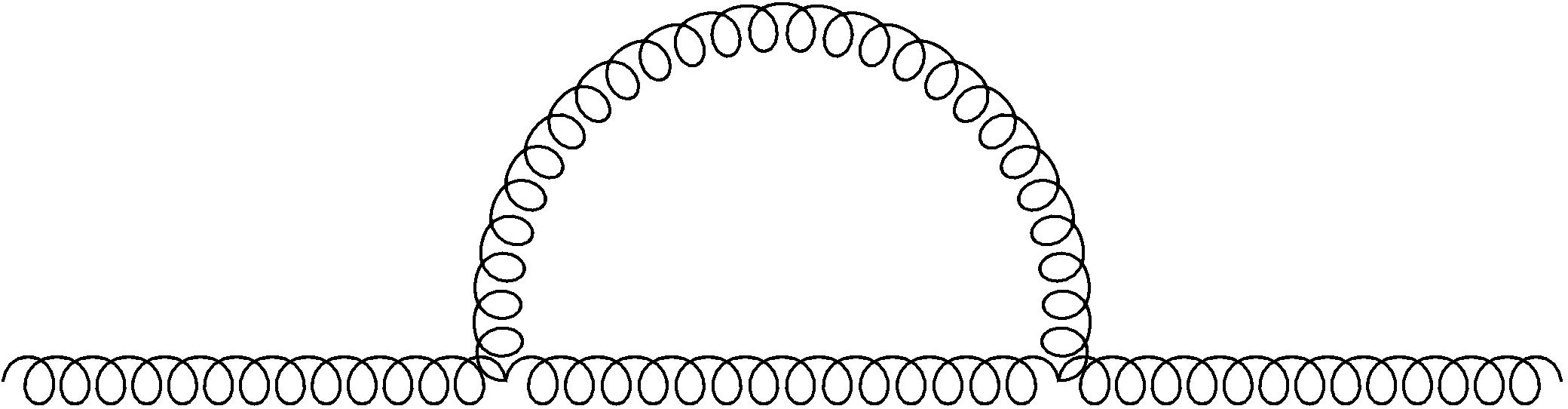}
    \includegraphics[scale=0.085]{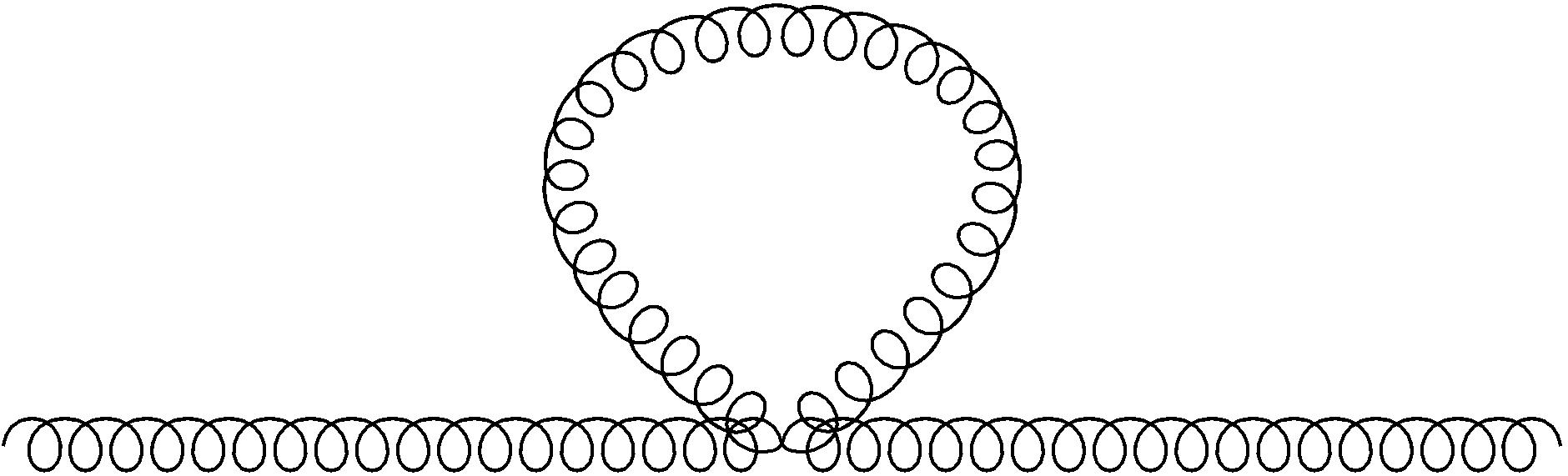}
    \includegraphics[scale=0.085]{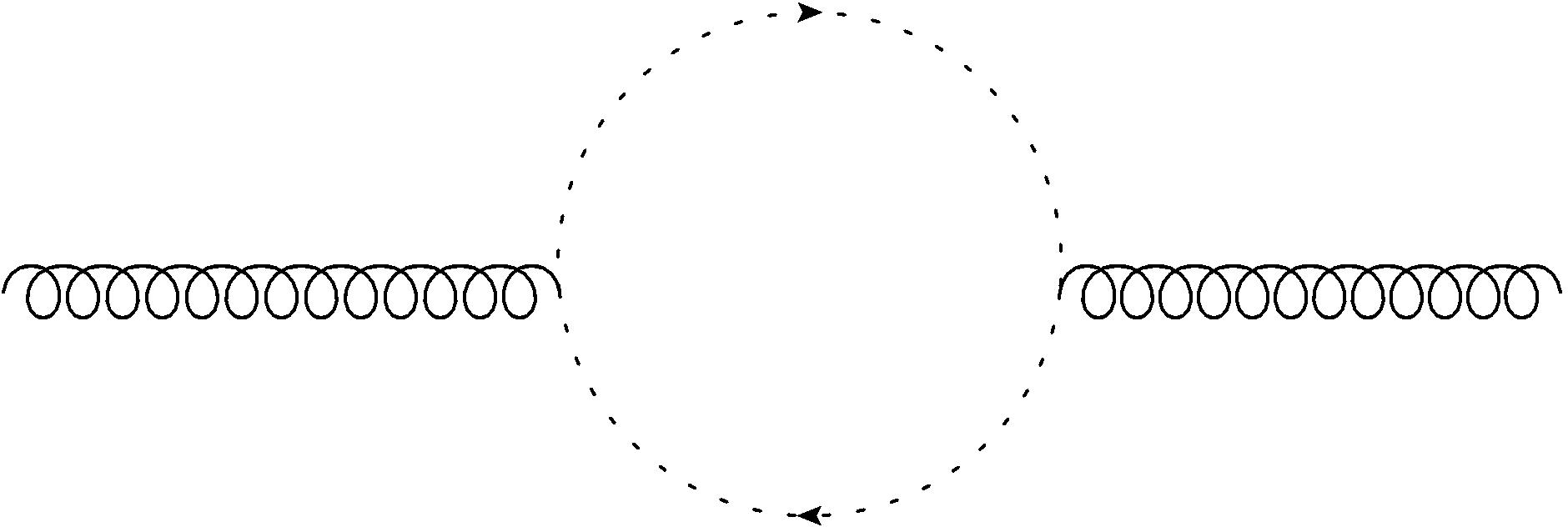}
    \caption{Sample Correction terms $SU(N)$ propagator diagram g10,g11,g12,g13,14 up to $O(1)$}
\end{figure}
Now it can be easily checked that for self interaction of the gauge field is suppressed by $O(1/M)$ thus we can drop them from the calculation,
\begin{equation}
    \pi_W(p)=-\frac{\Gamma(1/2)}{2(4\pi)^{3/2}p}\frac{\sqrt{\pi}\Gamma(1/2)}{2 \Gamma(2)}
\end{equation}

We can now define the exact propagator as double line,
\begin{figure}[h]
    \centering
    \includegraphics[scale=0.09]{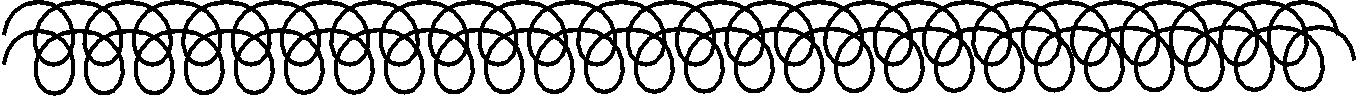}
    \caption{Corrected $SU(N)$ propagator upto $O(1/M)$}
\end{figure}
Now, we want to collect the correction to the $\psi$ propagator upto $O(1/M)$,
\begin{figure}[h]
    \includegraphics[scale=0.09]{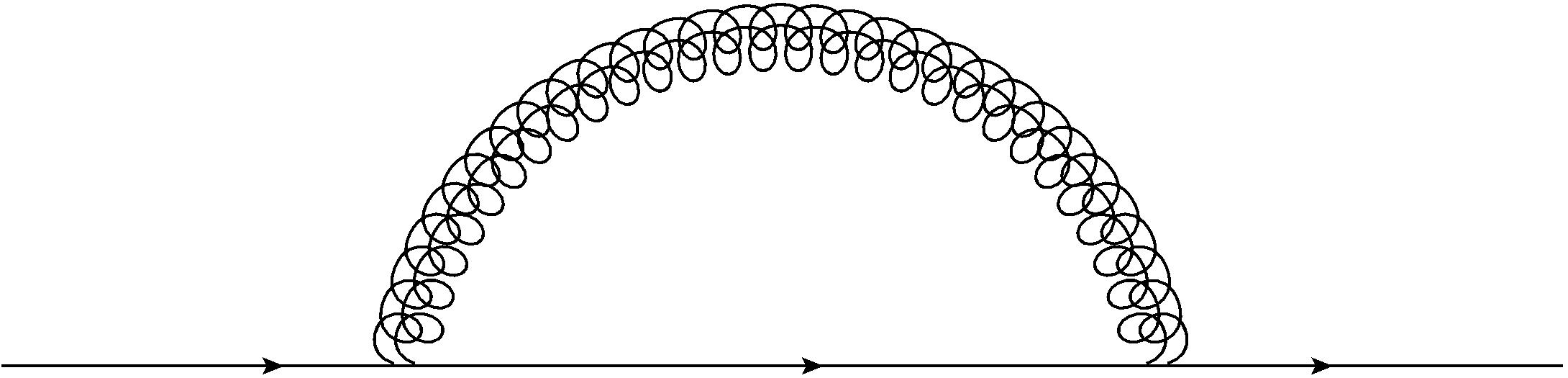}
    \includegraphics[scale=0.09]{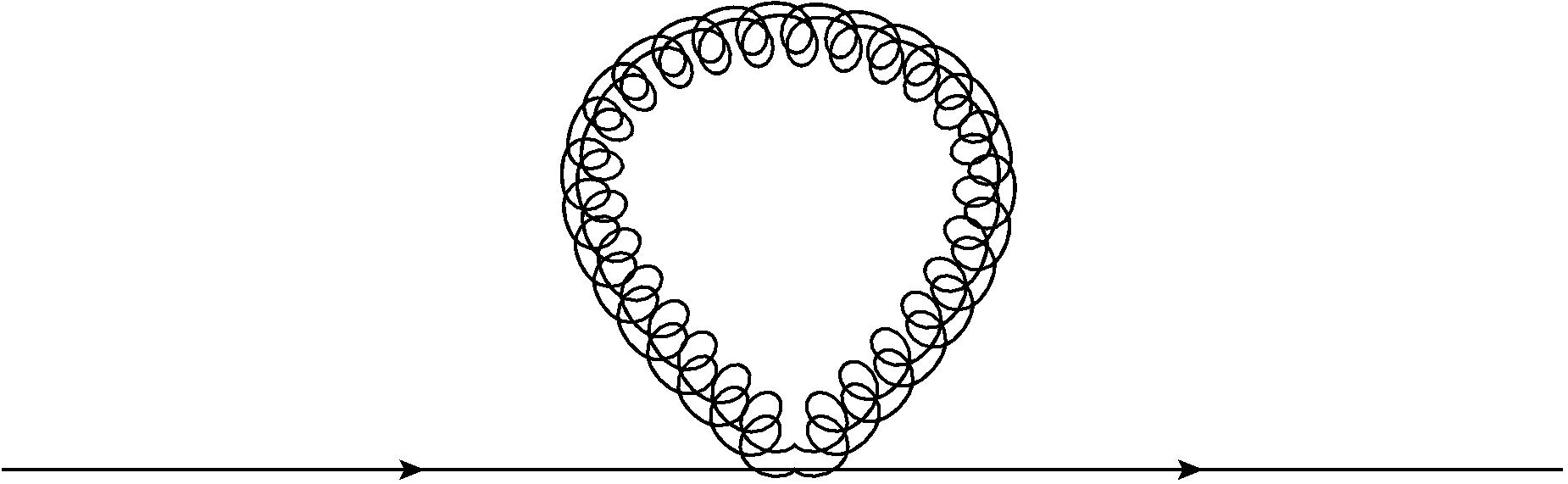}
    \caption{$SU(N)$ contribution to the propagator correction upto $O(1/M)$ diagram g16,g17}
\end{figure}
This calculation is exactly like $U(1)$ case which gives,
\begin{equation}
    \eta_W=\frac{64(N^2-1)}{3\pi^2 N M}
\end{equation}
Which gives,
\begin{equation}
    \boxed{\eta_{\text{total}}=-\frac{1}{N M}\left[ 2.0264+2.1615(N^2-1) \right]}
\end{equation}
\subsection{calculating $\gamma$ }
Next as in \cite{ma} we will switch on the mass. Then the fundamental bubble becomes,
\begin{multline}
    \pi(a,p)=\int \frac{d^3 l}{(2\pi)^3}\frac{1}{(l^2+a)((l+p)^2+a)}\\
    =\frac{\tan^{-1}\left(p/2\sqrt{a}\right)}{4\pi p}
\end{multline}
Now, to calculate $\gamma$ we have above the critical temperature(massive),
\begin{equation}
    G(0)\sim (a_o-a_{0c})^{-\gamma}
\end{equation}
Thus the leading diagram contributing to this is,
\begin{figure}[h]
    \centering
    \includegraphics[scale=0.09]{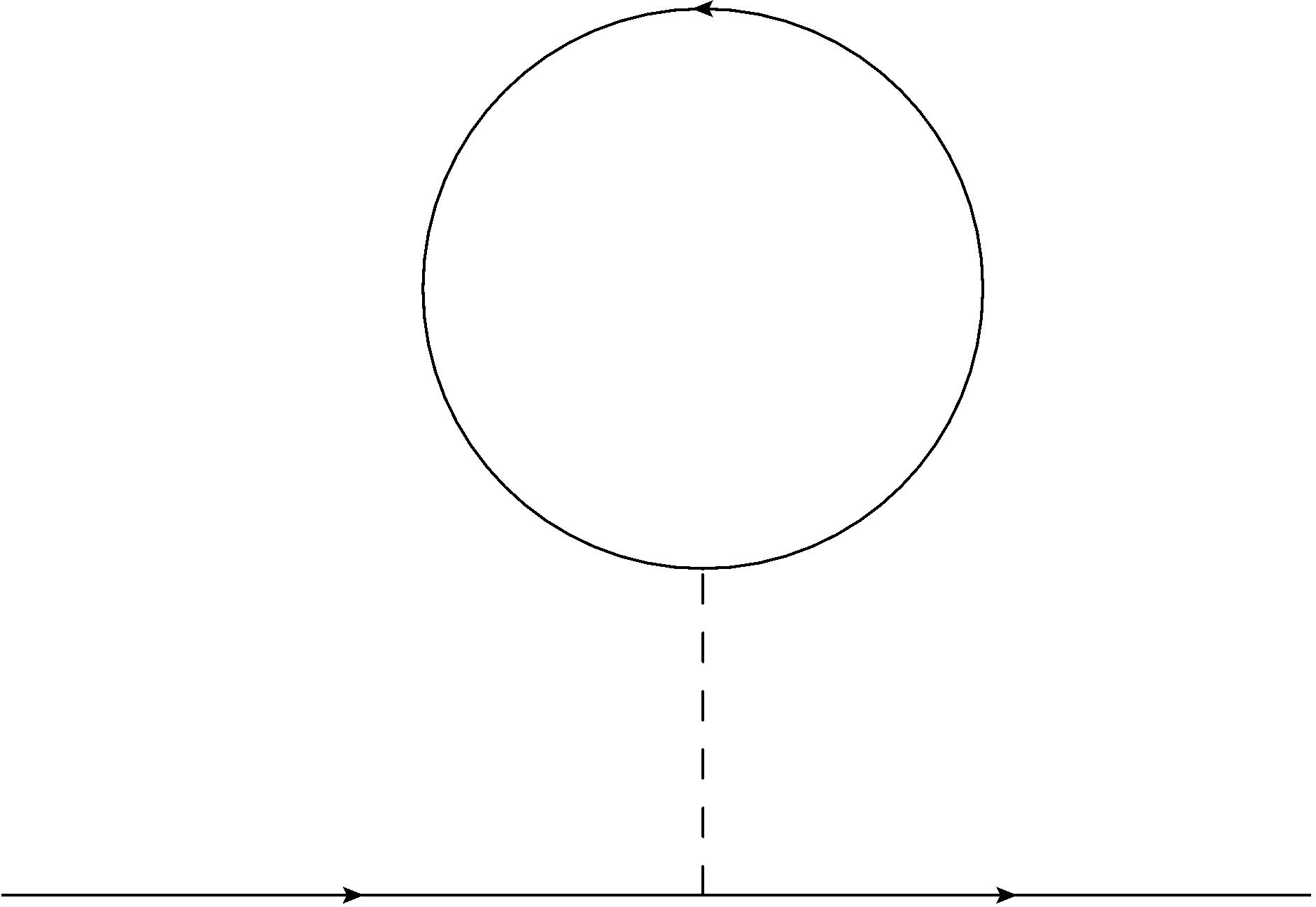}
    \caption{Leading diagram contributing to $\gamma$ diagram g18}
\end{figure}
And this contribution is,
\begin{multline}
    \Sigma_a(a)-\Sigma_a(0)=N M b\int\frac{d^3 l}{(2\pi)^3}\left[\frac{1}{(l^2+a)}-\frac{1}{l^2}\right]\\
    =-\frac{N M b a^{1/2}}{4\pi}
\end{multline}
Thus from leading calculation we get,
\begin{equation}
    \gamma=2
\end{equation}
Thus next order the diagram g19 does not contribute and can be shown easily.
\begin{figure}[h]
    \centering
    \includegraphics[scale=0.09]{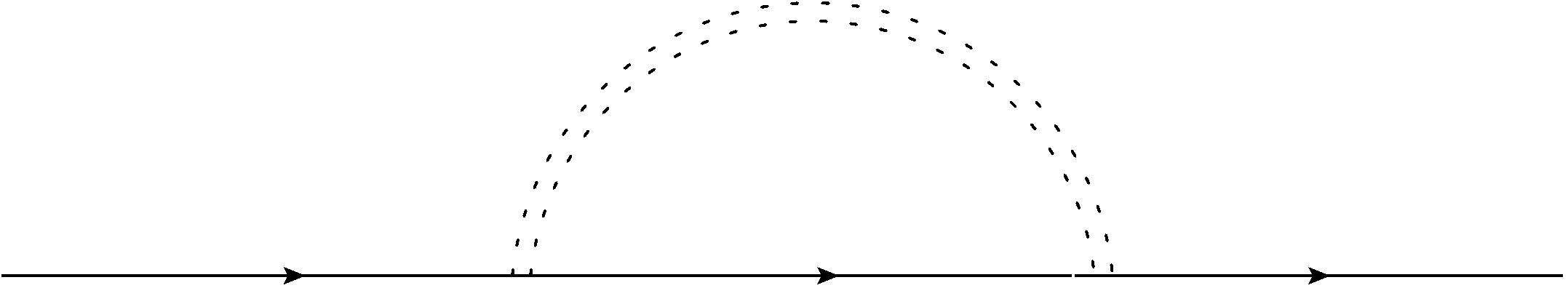}
    \caption{diagram g19 does not contribute directly to $\gamma$}
\end{figure}
The diagram g20 will contribute to the calculation of $\gamma$
\begin{figure}[h]
    \centering
    \includegraphics[scale=0.09]{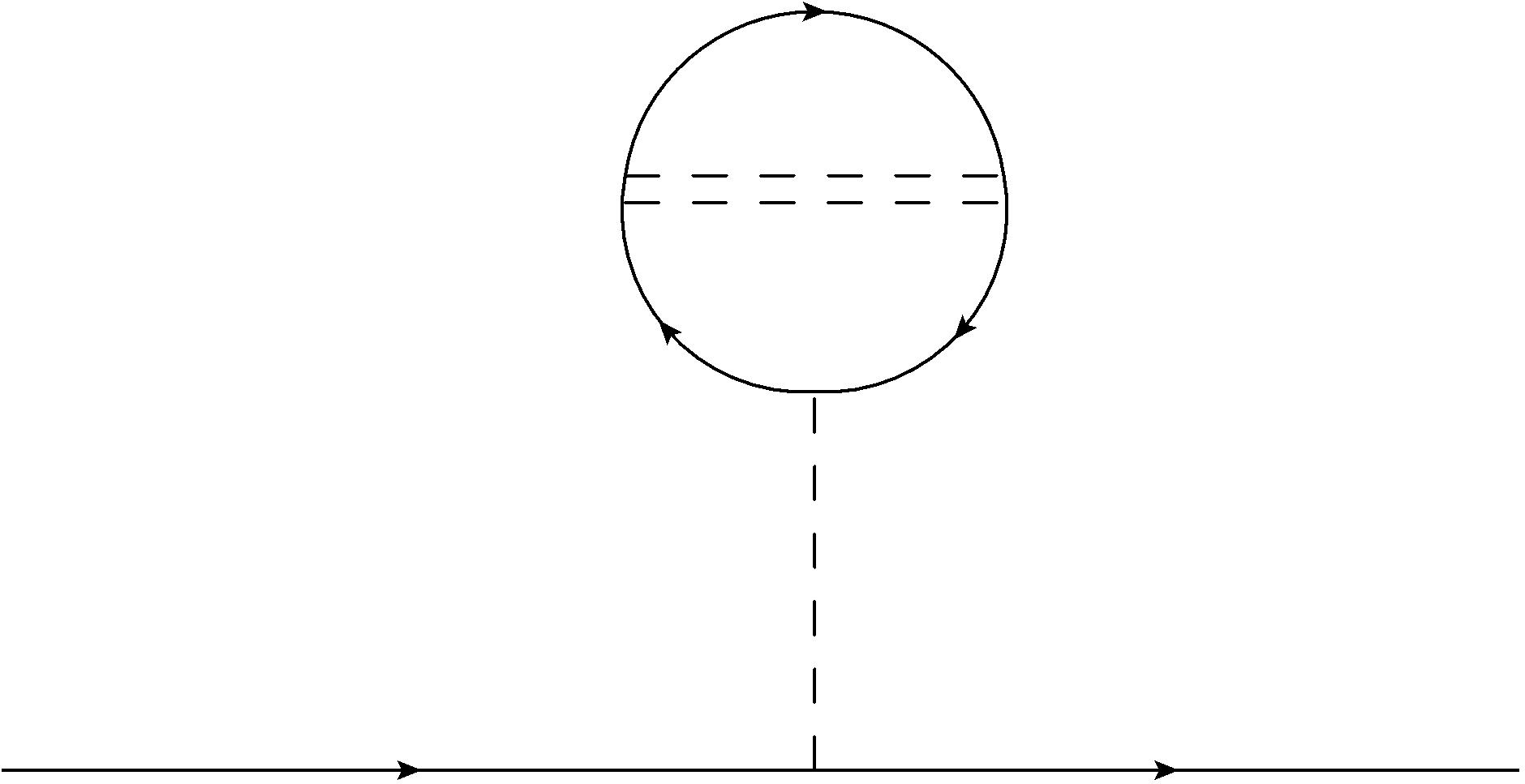}
    \caption{diagram g20 contributes to $\gamma$}
\end{figure}
And that gives,
\begin{multline}
    \Sigma_c(a,l)-\Sigma_c(a,0)\\
    =\int\frac{d^3 q}{(2\pi)^3}\frac{4\pi q}{N M\tan^{-1}(q/2\sqrt{a})}\left[ \frac{1}{(l+q)^2+a}-\frac{1}{q^2+a} \right]
\end{multline}
\begin{multline}
    \Sigma_b(a)=NMb\int\frac{d^3 l}{(2\pi)^3}\frac{1}{(l^2+a)^2}\left[ \Sigma_c(a,l)-\Sigma_c(a,0) \right]\\
    =\frac{ba^{1/2}}{8\pi^2}\int^{1/a}_0\frac{z dz}{\tan^{-1}(\sqrt{y}/2)}\left[\frac{1}{4+z}-\frac{1}{1+z}\right]
\end{multline}

All we need from this is to find the $\log[a]$ term of that integral. Defining the intregral to be $L$ we get,
\begin{equation}
    \frac{d L}{da}=\frac{3}{(a+5a^24a^3)\tan^{-1}(1/2\sqrt{a})}
\end{equation}
Now expanding and integrating back we get,
\begin{equation}
    \frac{1}{\gamma}=3/2-1+\frac{3}{\pi^2 NM}
\end{equation}
\begin{equation}
        \Rightarrow \gamma=2-\frac{12}{\pi^2NM}
\end{equation}
Now we want to introduce $U(1)$ gauge field then the fundamental bubble is,
\begin{figure}[h]
    \includegraphics[scale=0.09]{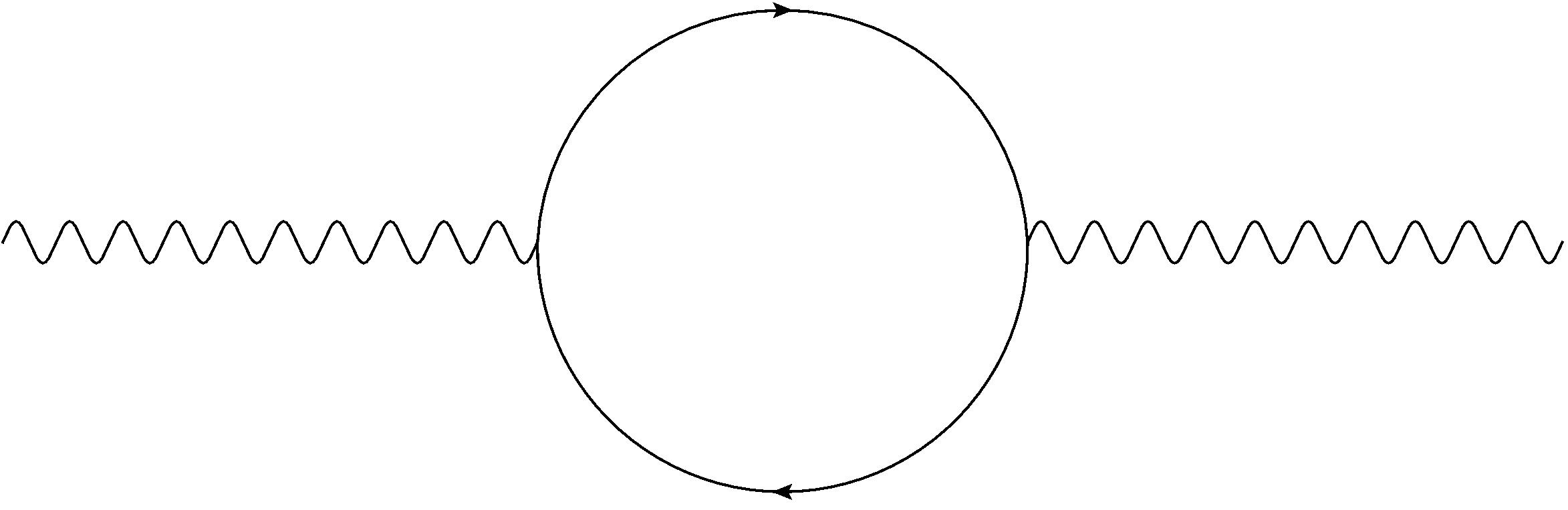}
    \includegraphics[scale=0.09]{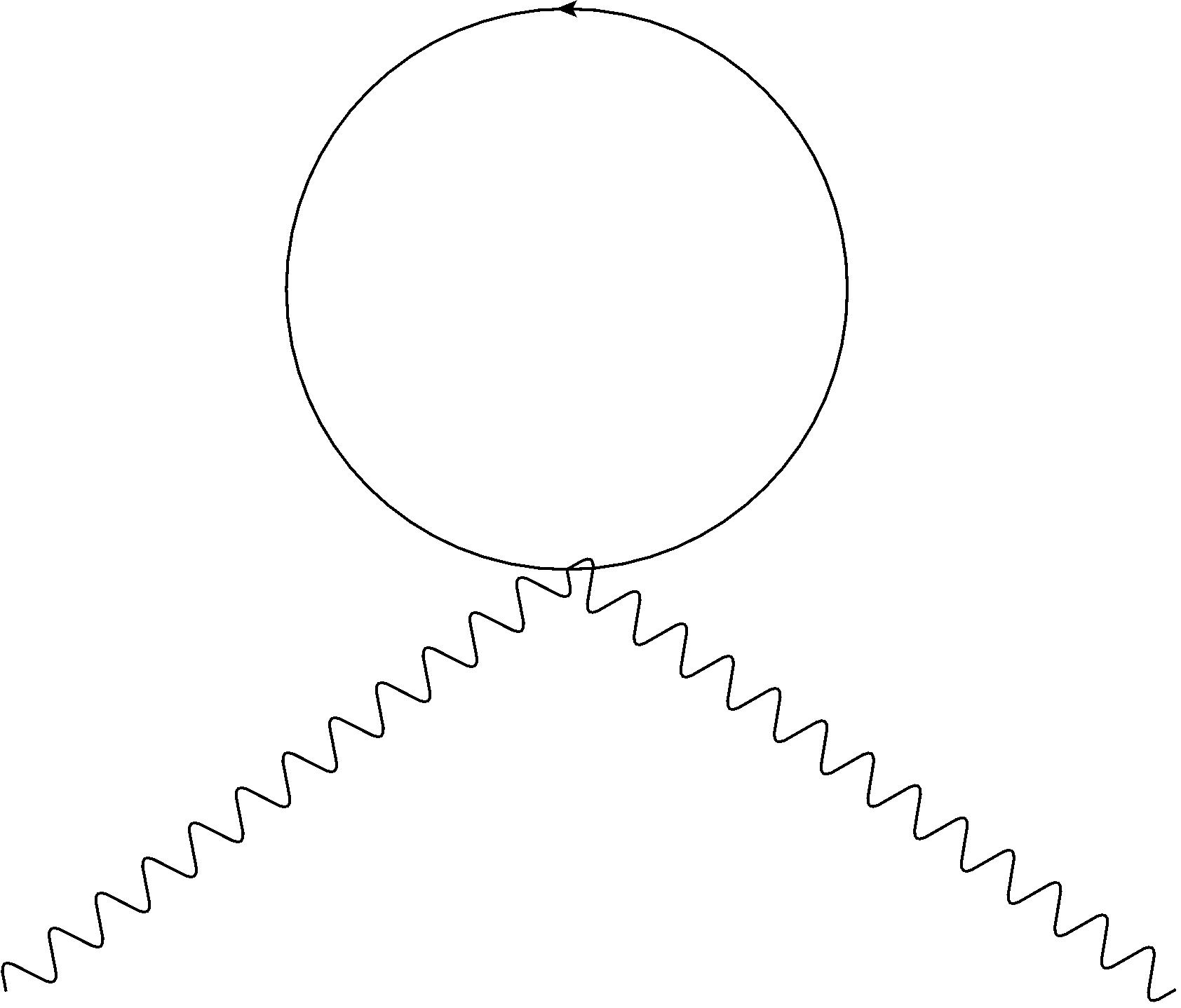}
    \caption{Fundamental bubble for the $U(1)$ propagator correction diagram g21,g22}
\end{figure}
Now,
\begin{multline}
    digram~g21+g22\\
    =NMy^2\int\frac{d^3 l}{(2\pi)^3}\left[ \frac{(p+2l)_\mu(p+2l)_\nu}{(l^2+a)((l+p)^2+a)}-\frac{2\delta_{\mu\nu}}{(l^2+a)} \right]\\
    \text{Now using Feynmann trick,}\\
    =-\frac{NMy^2}{8\pi p}(p^2\delta_{\mu\nu}-p_\mu p_\nu)\int_{-1/2}^{1/2}dx 4x^2\left[ (1/4-x^2)p^2+a \right]\\
    =\frac{NMy^2(p^2\delta_{\mu\nu}-p_\mu p_\nu)}{8\pi p}\left[ \frac{2\sqrt{a}}{p} -\left( 1+\frac{4a}{p^2} \right)\tan^{-1}\left(\frac{p}{2\sqrt{a}}\right) \right]
\end{multline}
Thus,
\begin{equation}
    \pi(a,p)=\frac{1}{8\pi p}\left[ \frac{2\sqrt{a}}{p} -\left( 1+\frac{4a}{p^2} \right)\tan^{-1}\left(\frac{p}{2\sqrt{a}}\right) \right]
\end{equation}
Now, similar to the case of the Hubbard-Stratonovich field the digrams that will contribute are diagram g23 and g24.
\begin{figure}[h]
    \includegraphics[scale=0.09]{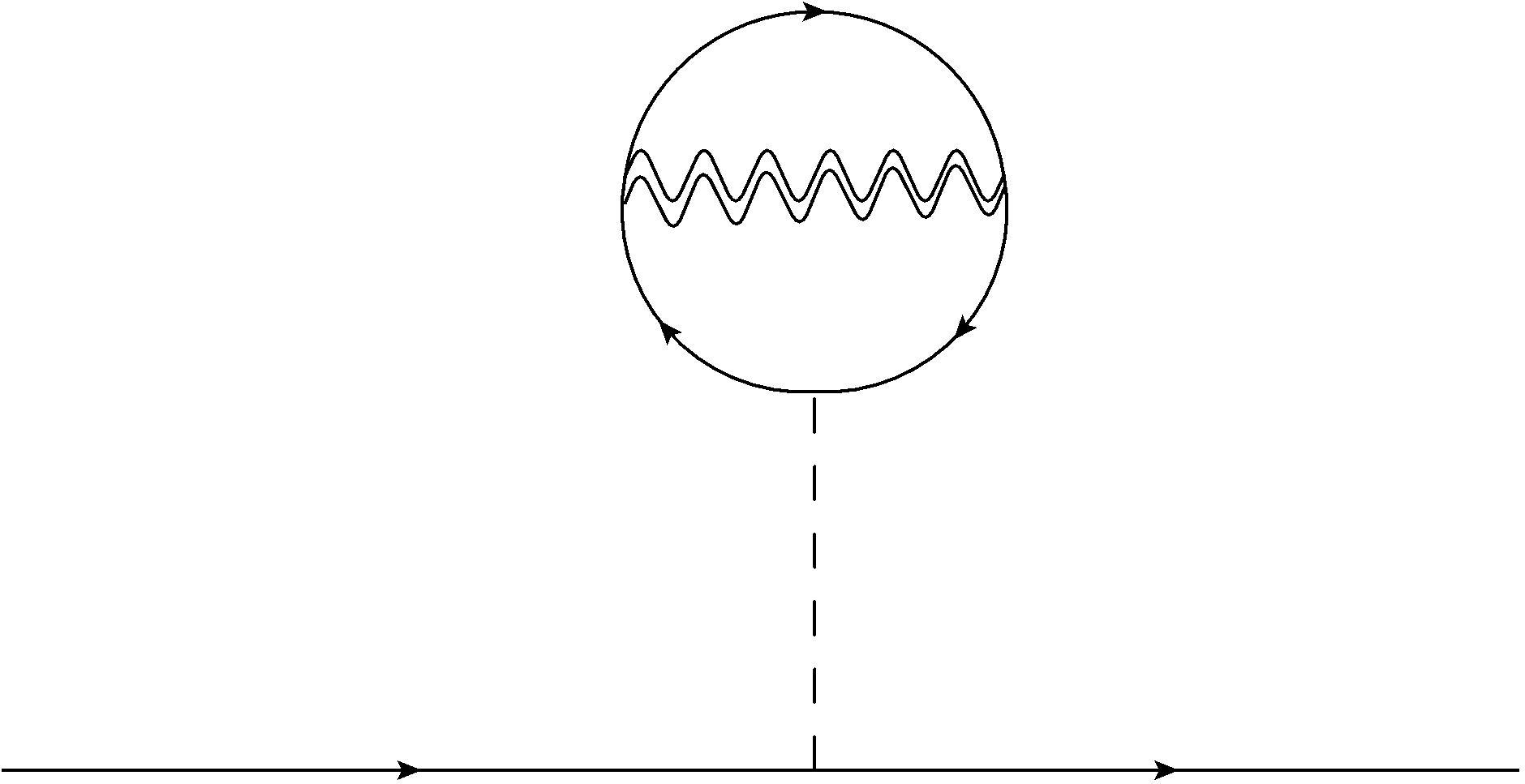}
    \includegraphics[scale=0.09]{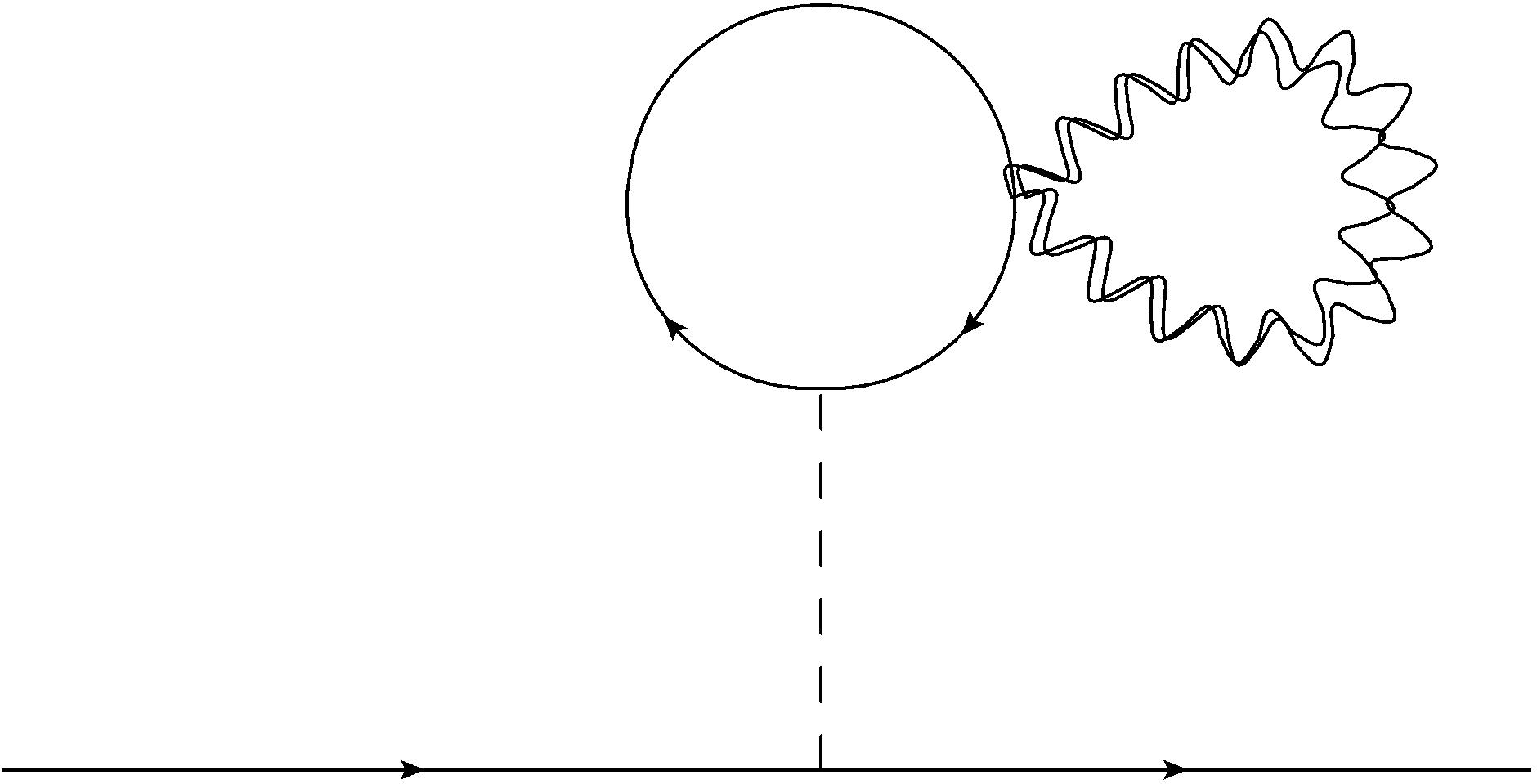}
    \caption{diagram g23,g24 contributes to $\gamma$}
\end{figure}
Now from this one can calculate,
\begin{multline}
    \Sigma_b=NMb\int \frac{d^3 l}{(2\pi)^3}\frac{1}{(l^2+a)^2}\left[ \Sigma_c(a,l)-\Sigma_c(a,0) \right]
\end{multline}
Where $\Sigma_c(a,l)$ is defined as sum of the diagram g25 and g26.
\begin{figure}[h]
    \includegraphics[scale=0.09]{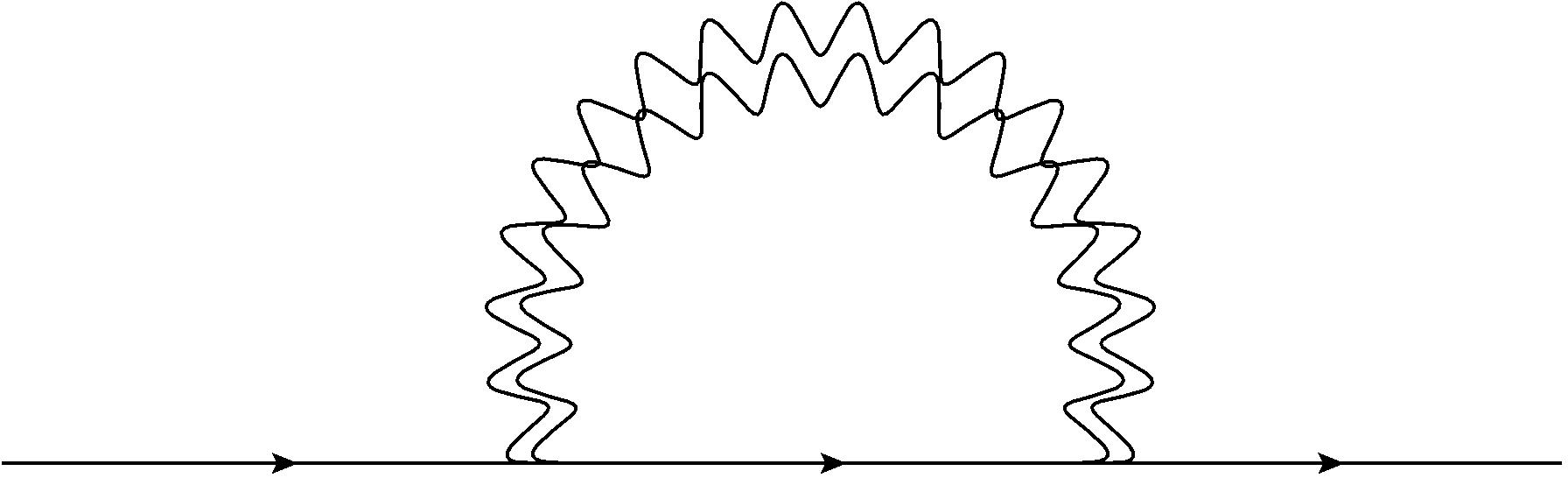}
    \includegraphics[scale=0.09]{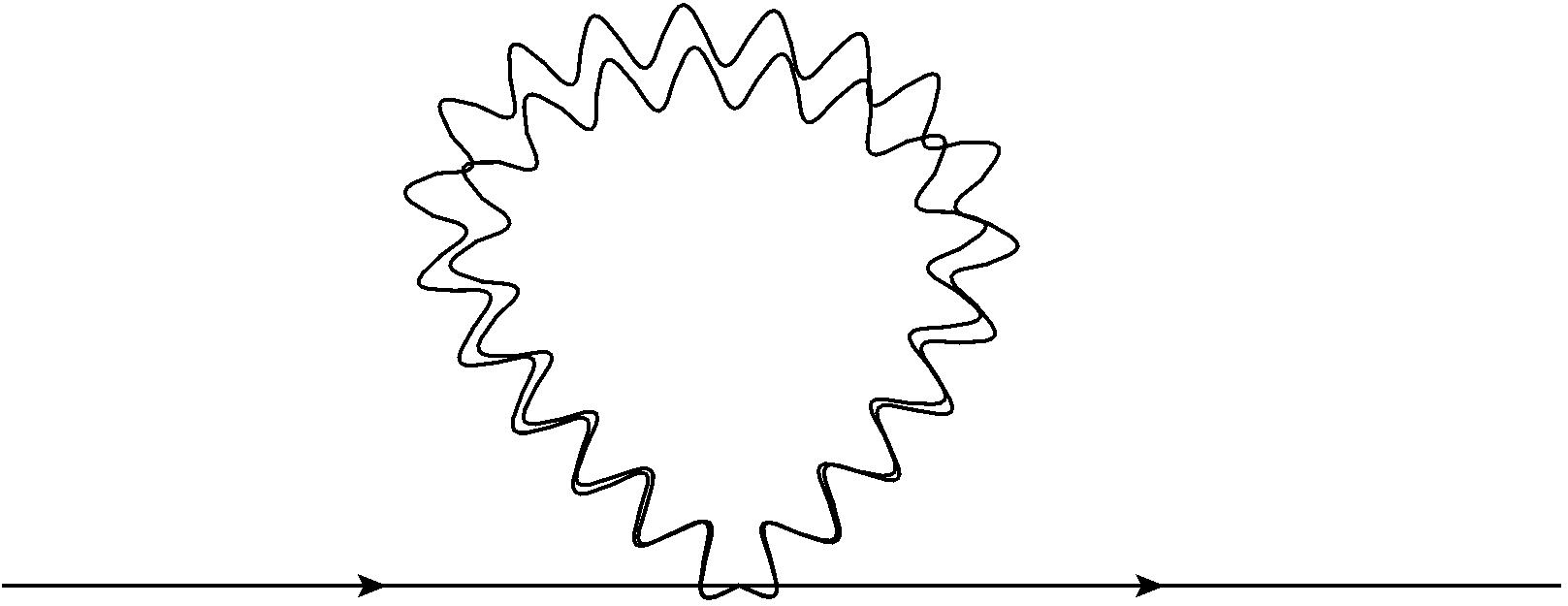}
    \caption{Diagram g25,g26 defining $\Sigma_c(a,l)$}
\end{figure}
This gives,
\begin{equation}
    \Sigma_b(a)=\frac{ba^{1/2}}{\pi^2}\int_0^{1/2}\frac{\tan^{-1}(\sqrt{z}/2)dz}{\sqrt{z}\left[ -2/\sqrt{z}+(1+4/z)\tan^{-1}(\sqrt{z}/2) \right]}
\end{equation}
This gives,
\begin{equation}
    1/\gamma=3/2-1+\frac{3}{\pi^2N M}+\frac{16}{\pi^2N M}=1/2+19/(\pi^2NM)
\end{equation}
\begin{equation}
    \Rightarrow \gamma=2-\frac{76}{\pi^2N M}
\end{equation}
In case of $SU(N)$ gauge field the fundamental bubble comes from diagram g27,g28
\begin{figure}[h]
    \includegraphics[scale=0.09]{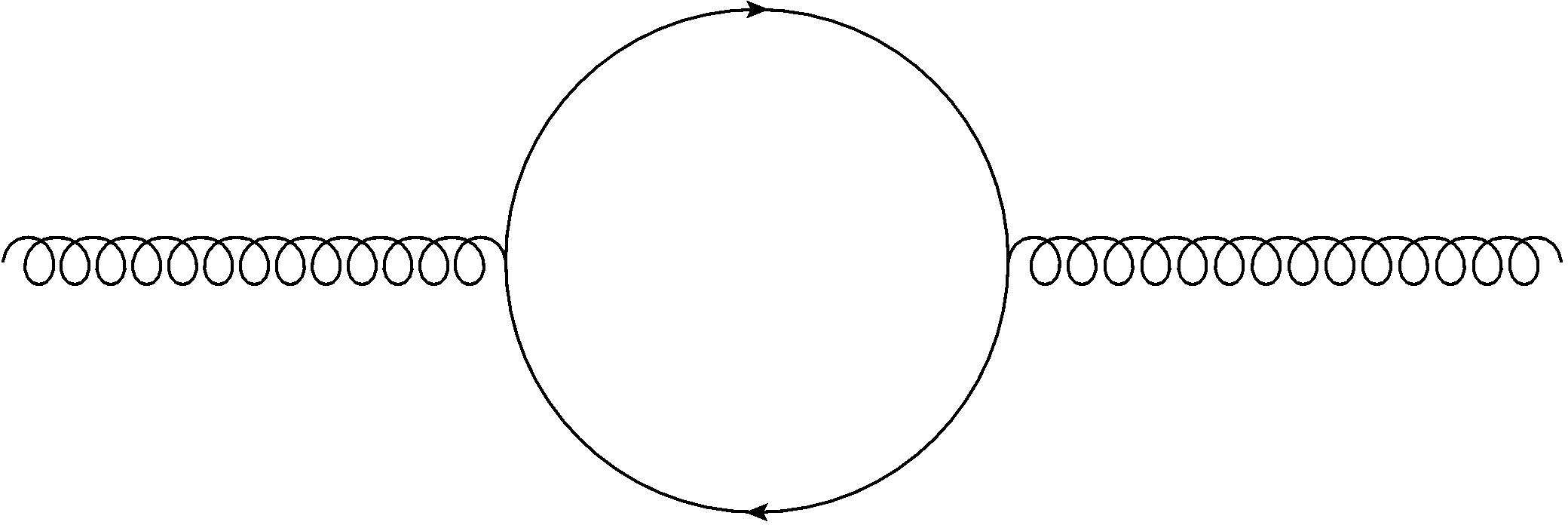}
    \includegraphics[scale=0.09]{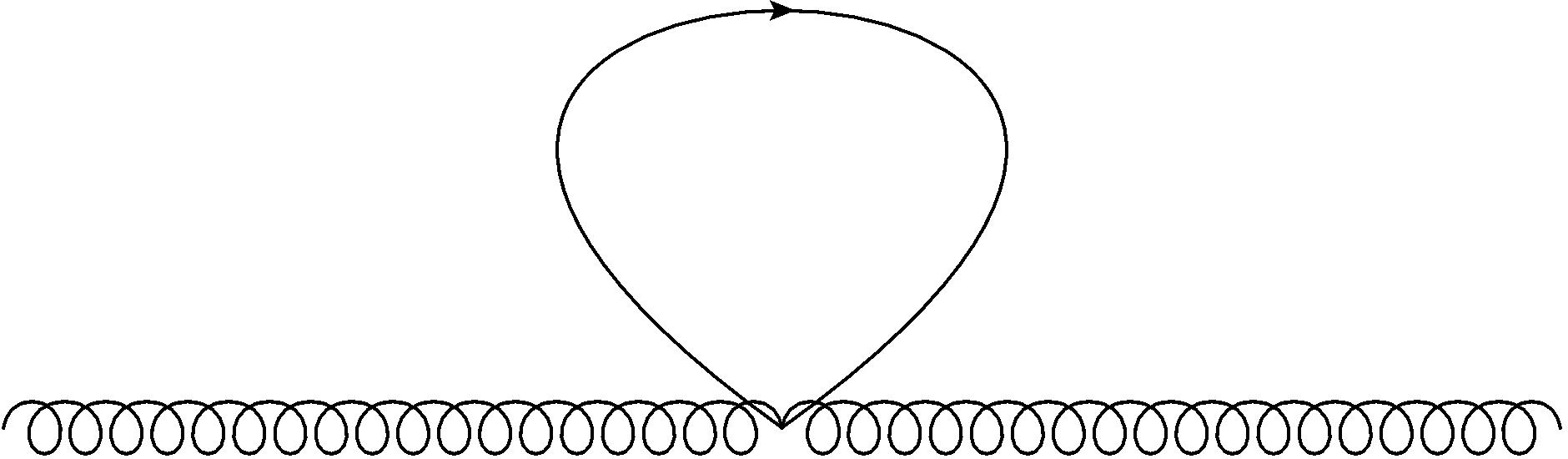}
    \caption{Fundamental bubble for the $SU(N)$ propagator correction diagram g27,g28}
\end{figure}
Then calculation similar to $U(1)$ case gives.
\begin{equation}
    \pi_W(a,p)=-\frac{1}{16\pi p}\left[ -\frac{2\sqrt{a}}{p} +\left( 1+\frac{4a}{p} \right)\tan^{-1}\left(\frac{p}{2\sqrt{a}}\right) \right]
\end{equation}
Then diagram g31,g32 gives the contribution to $\gamma$,
\begin{figure}
    \includegraphics[scale=0.09]{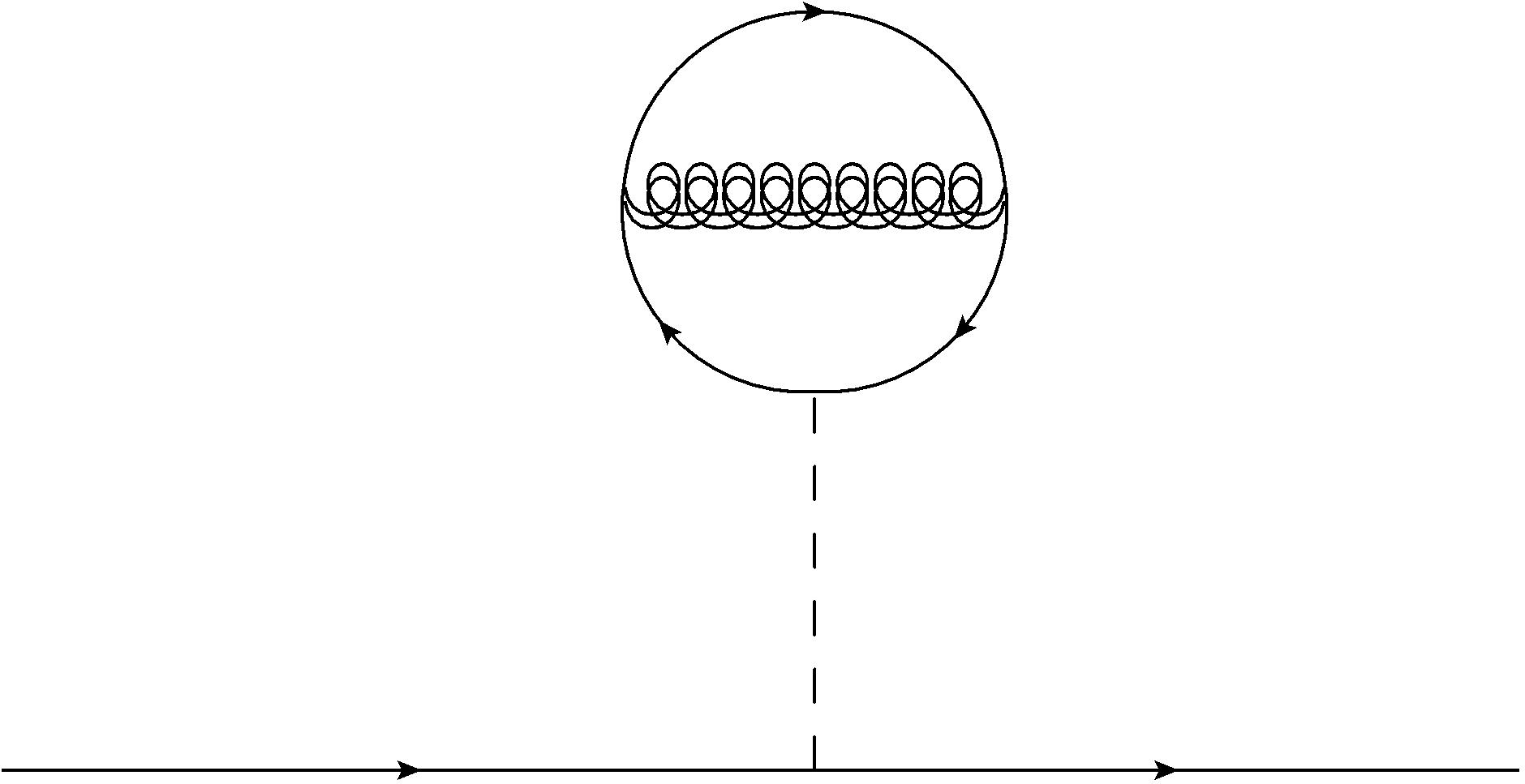}
    \includegraphics[scale=0.09]{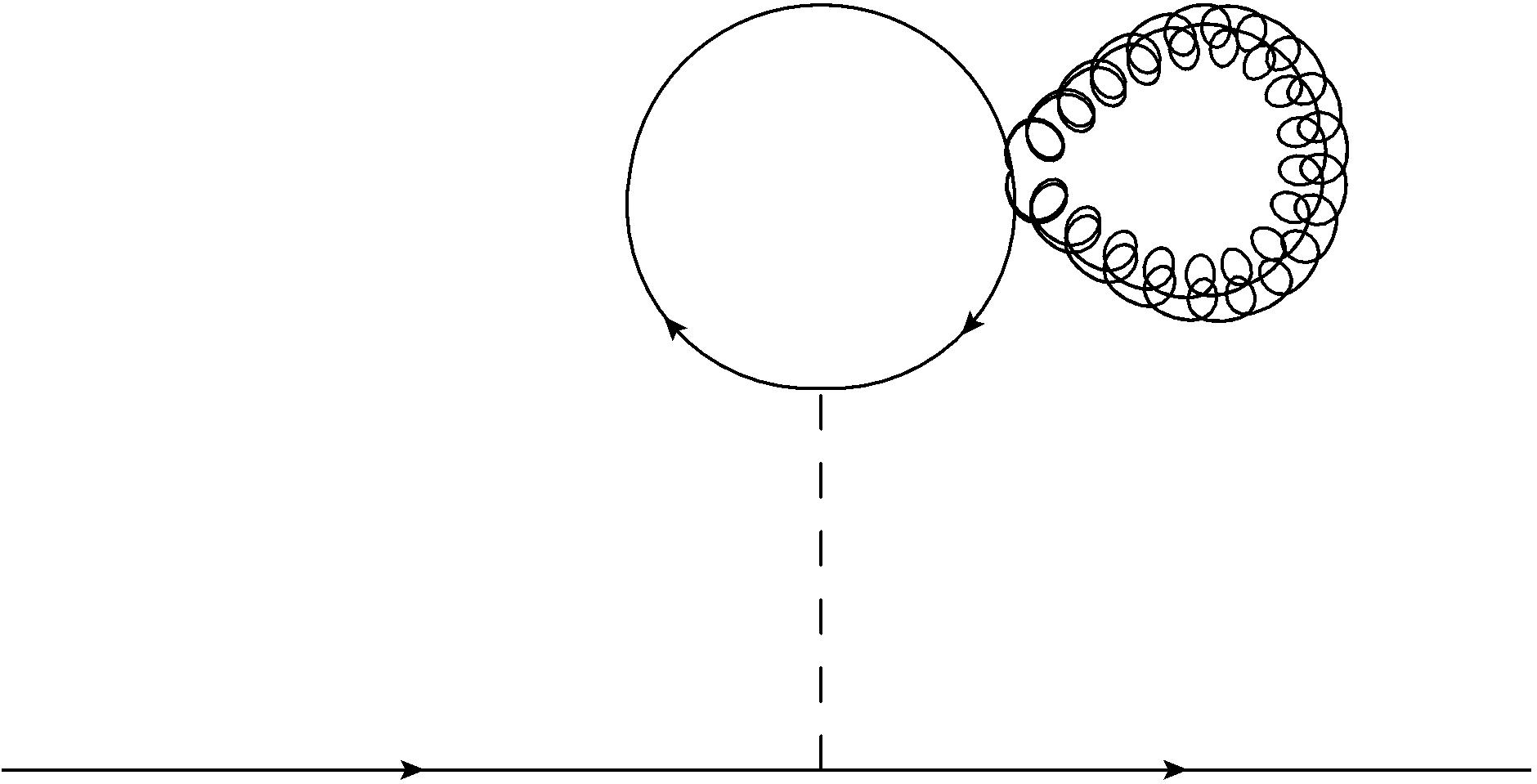}
    \caption{diagram g31,g32 contributes to $\gamma$}
\end{figure}
Now,
\begin{equation}
    \Sigma_b(a)=N M b\int\frac{d^3 l}{(2\pi)^3}\frac{1}{(l^2+a)^2}\left[ \Sigma_c(a,l)-\Sigma_c(a,0) \right]]
\end{equation}
Where $\Sigma_c(a,l)$ is diagram g29+g30.
\begin{figure}[H]
    \includegraphics[scale=0.09]{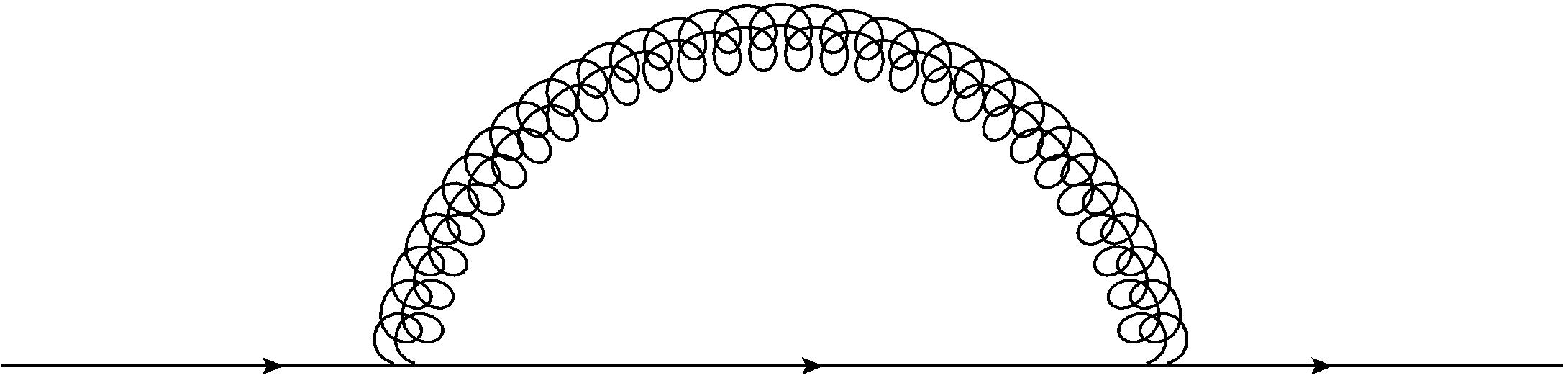}
    \includegraphics[scale=0.09]{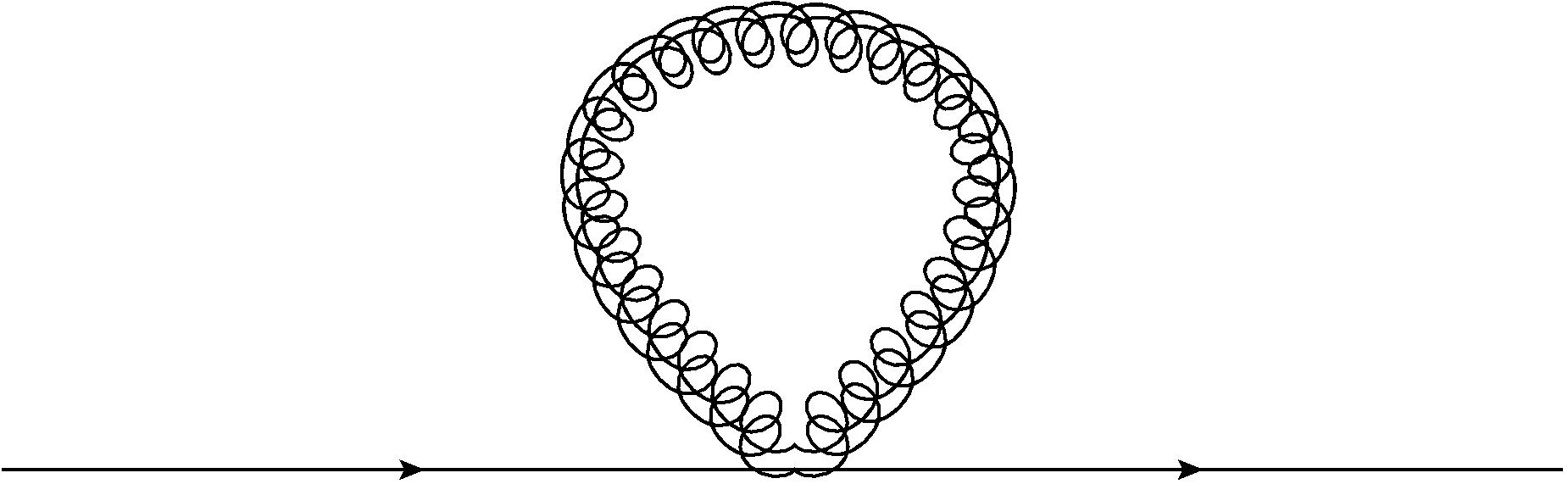}
    \caption{Diagram g29,g30 defining $\Sigma_c(a,l)$}
\end{figure}
Thus similar to $U(1)$ case,
\begin{equation}
    \frac{1}{\gamma}=\frac{1}{2}+\frac{19}{\pi^2NM}+\frac{16(N^2-1)}{\pi^2NM}
\end{equation}
\begin{equation}
    \Rightarrow \boxed{\gamma_\text{total}=2\left[1-\frac{1}{N M}(3.8502+3.2423(N^2-1))\right]}
\end{equation}

\subsection{Calculation of $\nu$}
From scaling law we have,
\begin{equation}
    \nu=\frac{\gamma}{2-\eta}
\end{equation}
Thus here we get,
\begin{equation}
    \boxed{\nu=1-\frac{4.86}{NM}-\frac{4.32(N^2-1)}{NM}}
\end{equation}

\begin{thebibliography}{99}
{
        
    \bibitem{sachdev} 
        N. Read and Subir Sachdev, Phys. rev. B, Volume 42, issue 7, 1990
    
    \bibitem{senthil}
        Tarun Grover and T. Senthil, Phys. Rev. Lett., Volume 98, issue 24, 2007
        
    \bibitem{senthil2}
        Tarun Grover and T. Senthil, Phys. Rev. Lett., Volume 107, issue 5, 2011
    
    \bibitem{lammert}
        Paul E. Lammert, Daniel S. Rokhsar, and John Toner, Phys. Rev. E., Volume 52, issue 2, 1995
    
    \bibitem{lee}
        Sung-Sik Lee and Patrick A. Lee, Phys. Rev. B, Volume 72, issue 23, 2005
    
    \bibitem{huo}
        Qiu-Hong Huo and Yunguo Jiang and Ru-Zhi Wang and Hui Yan, Europhysics Letters, Volume 101, issue 2, 2013
    
    \bibitem{blasone}
        Massimo Blasone, Petr Jizba and Giuseppe Vitiello, Quantum Field Theory and Its Macroscopic Manifestations, Imperial College Press, 2011
    
    \bibitem{gor} 
        L. P. GOR'KOV, SOVIET PHYSICS JETP, VOLUME 36(9), NUMBER 6, DECEMBER, 1959

    \bibitem{weinberg}
        Sidney Coleman and Erick Weinberg, Phys. Rev. D, Volume  7, issue 6, 1973
    
    \bibitem{HaLuMa}
        Halperine, Lubensky \& Ma, PRL, Volume 32, Number 6,1974
    
    \bibitem{arnold}
        Arnold, Peter and Yaffe, Laurence G., Phys. Rev. D, Volume 49, issue 6, 1994
        
    \bibitem{laine} 
        K. Kajantie, M. Laine, K. Rummukainen, and M. Shaposhnikov, Phys. Rev. Lett., volume 77,issue 14, 1996
    
    \bibitem{laine2} 
        K.Kajantie, M.Laine, K. Rummukainen, M. Shaposhnikov,Nuclear Physics B, Volume 493, Issues 1–2, 1997
    
    \bibitem{laine3} 
        K.Kajantie, M.Laine, M. Shaposhnikov, Nuclear Physics B, Volume 407, Issue 2, 1993

    \bibitem{fradkin} 
        Fradkin and Shenker, Phys. Rev. D 19, 3862, 1979
    
    \bibitem{arnold2} 
        Peter Arnold and David Wright, Phys. Rev. D, volume 55, issue 10, 1997
        
    \bibitem{auerbach}
        Arovas and Auerbach, Phys. Rev. B, Volume 38, issue 1, 1988
        
    \bibitem{auerbach2}
        Auerbach and Arovas, Phys. Rev. Lett., Volume 61, issue 5, 1988
    
    \bibitem{naoki} 
        Naoki Kawashima and Yuta Tanabe, Phys. Rev. Lett., volume 98, issue 5, 2007
    
    \bibitem{beach}  
        Beach, K. S. D. and Alet, Fabien and Mambrini, Matthieu and Capponi, Sylvain, Phys. Rev. B volume 80, issue 18,2009
            
    \bibitem{pes}
         Peskin \& Schroeder, An Introduction to Quantum Field Theory, Westernview Press, 2005
    
    \bibitem{sred}
         Mark Srednicki, Quantum Field Theory, Cambridge University Press, 2007
    
    \bibitem{igor}
        IGOR HERBUT, A Modern Approach to Critical Phenomena, Cambridge University Press, 2007
        
    \bibitem{kiom} 
        Michael Kiometzis, Hagen Kleinert, and Adriaan M. J. Schakel, Phys. Rev. Lett., Volume 73, issue 14, 1994
        
    \bibitem{ma}
        Shang-Keng Ma, Phys. Rev. A, Volume 7, issue 6, 1973
        
    \bibitem{sakhi} S. Sakhi, Phys. Rev. D, Volume 90, issue 4, 2014
    
     \bibitem{Moore}
        Moore, M. A. and Newman, T. J. and Bray, A. J. and Chin, S-K., Phys. Rev. B, volume 58, issue 2, 1998
}
\end{thebibliography}
\end{document}